\documentclass[aps,prd,reprint,amsmath,amssymb,showpacs,superscriptaddress]{revtex4-1}
\usepackage[colorlinks,linkcolor=blue,anchorcolor=blue,citecolor=blue,urlcolor=blue,breaklinks=true]{hyperref}
\usepackage{graphicx}
\usepackage{slashed}
\usepackage{epsfig}
\usepackage{amsbsy}
\usepackage{array}
\usepackage{changes}
\usepackage{amssymb}
\usepackage{setspace}
\usepackage{bm}
\usepackage{lipsum}
\usepackage{mathrsfs}
\usepackage{color}
\usepackage{graphicx}
\usepackage{subfigure}
\usepackage{mathrsfs}
\usepackage{ulem}
\usepackage{MnSymbol}
\usepackage[export]{adjustbox}

\begin{document}
\author{Yin-Zhen Xu}
\affiliation{Department of physics, Nanjing University, Nanjing 210093, China}

\author{Si-Xue Qin}
\email{sqin@cqu.edu.cn}
\affiliation{Department of Physics, Chongqing University, Chongqing 401331, China}
\affiliation{Chongqing Key Laboratory for Strongly-Coupled Physics, Chongqing 401331, China}

\author{Hong-Shi Zong}
\affiliation{Department of physics, Nanjing University, Nanjing 210093, China}
\affiliation{Department of physics, Anhui Normal University, Wuhu 241000, China}
\affiliation{Nanjing Proton Source Research and Design Center, Nanjing 210093, China}
\affiliation{Joint Center for Particle, Nuclear Physics and Cosmology, Nanjing 210093, China}

\date{\today}

\title{Chiral symmetry restoration and properties of Goldstone bosons at finite temperature}

\begin{abstract}
	We study chiral symmetry restoration by analyzing thermal properties of QCD's (pseudo-)Goldstone bosons, especially the pion. The meson properties are obtained from the spectral densities of mesonic imaginary-time correlation functions. To obtain the correlation functions, we solve the Dyson-Schwinger equations and the inhomogeneous Bethe-Salpeter equations in the leading symmetry-preserving rainbow-ladder approximation. In the chiral limit, the pion and its partner sigma degenerate at the critical temperature $T_c$. At $T \gtrsim T_c$, it is found that the pion rapidly dissociates, which signals deconfinement phase transition. Beyond the chiral limit, the pion dissociation temperature can be used to define the pseudo-critical temperature of chiral phase crossover, which is consistent with that obtained by the maximum point of the chiral susceptibility. The parallel analysis for kaon and pseudoscalar $s\bar{s}$ suggests that heavy mesons may survive above $T_c$.
	
\end{abstract}

\maketitle

\maketitle

\section{Introduction}

The quantum chromodynamics(QCD) as a fundamental theory describing strong interaction exhibits two fascinating nonperturbative features: confinement and dynamical chiral symmetry breaking (DCSB). Those features govern properties of QCD bound states--hadrons. At finite temperature, the heat bath affects the dynamics and thus properties of hadrons, e.g., hadron masses, decay constants, and etc. At high temperature, the hadron matter undergoes phase transitions: deconfinement and chiral restoration, and a new state of matter, i.e., quark-gluon plasma (QGP), may form \cite{Cabibbo:1975ig,Sarkar:2010zza}. Studies on QCD phase transition have been being both experimental and theoretical frontiers \cite{Fischer:2018sdj,Fukushima:2010bq,Roberts:2000aa,Ding:2015ona,Qin:2010nq,Gao:2020qsj,Braun:2020ada,Fu:2019hdw}.

At finite temperature, the thermal modifications may occur for hadron properties, e.g., mass, size, structure, and etc. Among these mesons, the pion $\pi$ is of particular importance. Since the pion has a twofold role: the Goldstone mode of DCSB and the quark-antiquark bound state embodying confinement, the temperature dependence of its properties is critical to understand the QCD phase transitions. Moreover, its chiral partner, i.e., scalar meson $\sigma$, plays also an important role. In vacuum, a sizable $\pi$-$\sigma$ mass splitting can be observed. But they tend to degenerate in heat bath, which may indicate the chiral symmetry restoration.

The properties of light mesons at finite temperature have attracted numerous attentions and been studied by many approaches in recent years, e.g., lattice QCD \cite{Brandt:2015sxa,Cheng:2010fe}, chiral perturbation theory (ChPT) \cite{Toublan:1997rr,Andersen:2012dz}, Nambu--Jona-Lasinio (NJL) model \cite{Bernard:1987ir,Sheng:2020hge}, Polyakov-loop-extended NJL model \cite{Wergieluk:2012gd,Blaschke:2017pvh}, and others \cite{Hatsuda:1987kg,Andersen:2006ys,Cui:2013zha,Bartz:2016ufc}. The QCD's Dyson-Schwinger equations (DSE), a nonperturbative and Poincar\'{e} covariant framework which is capable of simultaneously describing confinement and DCSB, has been widely used to study hadron properties in both vacuum and medium\cite{Roberts:2007ji,Roberts:1994dr,Xu:2019ilh,Maris:2003vk,Fischer:2018sdj,Maris:2000ig,Gao:2020hwo,Qin:2013aaa,Chen:2020afc}. Mesons, the two-body bound states, correspond to the poles of the scattering matrices between quarks and antiquarks at zero temperature. Thus, the bound-state equation, i.e., homogenous Bethe-Salpeter equation (BSE), is defined by the on-shell condition. However, at finite temperature, the situation is different. Mesons in heat bath may dissociate \cite{Yagi:2005yb}. In such case, the poles may be smeared and become peaks with finite widths, and thus the homogenous BSE cannot be well defined any more. Moreover, in order to solve the homogenous BSE, one needs the analytically continued quark propagators as input. The analytical continuation is difficult for the quark propagators defined on the Matsubara frequencies in the imaginary-time formalism of thermal field theory.

In this work, we analyze the spectral functions of mesonic correlation functions to study properties of pseudoscalar and scalar mesons at finite temperature by following Refs. \cite{Chen:2020afc,Qin:2013aaa}. The imaginary-time correlation functions are computed with the solutions of the quark gap equation and the inhomogeneous BSE. For solving them, we adopt the so-called rainbow-ladder (RL) truncation, which approximates the quark-gluon vertex as the bare one and the BSE kernel as the one-gluon-exchange form. The RL truncation is the leading symmetry-preserving approximation, which can realize the twofold role of pion, i.e., the quark-antiquark bound state and the Goldstone boson \cite{Delbourgo:1979me,Bender:1996bb,Munczek:1994zz}.

In Section \ref{sec:gap}, we describe the gap equation of the quark propagator at finite temperature and study the chiral phase transitions. In Section \ref{sec:bse}, we analyze the relationship between thermal properties of $\pi$ and $\sigma$ and chiral symmetry restoration. At last, we extend the analysis by studying the cases of kaon and $s\bar{s}$ pseudoscalar meson. In Section \ref{sec:summary}, we present a brief summary and perspectives.

\section{Quark gap equation and chiral symmetry restoration} \label{sec:gap}
The quark gap equation at finite temperature has been described in Refs. \cite{Roberts:2000aa,Fischer:2018sdj}, which reads
\begin{eqnarray}
&&S\left({\omega}_{n}, \vec{p}\right)^{-1} = Z_{2}\left(i \vec{\gamma} \cdot \vec{p}+i \gamma_{4} {\omega}_{n}+Z_{m} m\right) + \Sigma({\omega}_{n}, \vec{p})\,,\quad
\end{eqnarray}
with the self-energy in the rainbow approximation
\begin{eqnarray}
\Sigma\left({\omega}_{n}, \vec{p}\right) = \frac{4}{3} Z_{1} T \sumint_l \frac{d^{3} p}{(2 \pi)^{3}} g^{2} D_{\mu \nu}\left(k_{\Omega}\right) \gamma_{\mu} S\left({\omega}_{l}, \vec{q}\right) \gamma_{\nu} \,,\quad
\end{eqnarray}
where $ {\omega}_{l}=(2 l+1) \pi T$ are the fermionic Matsubara frequencies; $m$ is the current quark mass which defines the chiral limit for $m = 0$; $Z_{1,2,m}$ are the quark-gluon vertex, quark wave function, and mass renormalization constants, respectively; and $\sumint_l$ represents the summation of Matsubara frequencies and the three-dimensional integral. $ D_{\mu \nu}\left(k_{\Omega}\right)$ is the dressed-gluon propagator which has the form:
\begin{align}
g^{2} D_{\mu \nu}\left(k_{\Omega}\right)=P_{\mu \nu}^{T} \mathcal{D}\left(k_{\Omega}^{2}\right)+P_{\mu \nu}^{L}\mathcal{D}\left(k_{\Omega}^{2}+m_{g}^{2}\right),
\end{align}
where the gluon momentum $k_{\Omega}:=\left({\omega}_{n}-{\omega}_{l}, \vec{p}-\vec{q}\right)$, the gluon Debye mass $m_g^2=(16 / 5) \pi^{2} T^{2}$, and $\mathcal{D}$ stands for the interaction kernel. Note that the temperature breaks the Euclidean $O(4)$ symmetry, and thus the gluon tensor structure breaks into  transverse and longitudinal parts, accordingly,
\begin{eqnarray}
P_{\mu \nu}^{T} &:=&\left\{\begin{array}{ll}{0,~\quad\qquad\qquad} {\mu \text { and/or } \nu=4} \\ {\delta_{i j}-\frac{k_{i} k_{j}}{k^{2}},\quad\quad} {\mu, \nu=i, j=1,2,3}\end{array}\right.,\\
P_{\mu \nu}^{L} &:=&\delta_{\mu \nu}-\frac{k_{\mu} k_{\nu}}{k^{2}} - P_{\mu \nu}^{T}.
\end{eqnarray}

The interaction kernel must be specified. In general, phenomenological models are adopted \cite{Maris:2003vk, Qin:2011dd}. According to the modern DSE and lattice QCD, the gluon mass scale is the most important feature of the interaction kernel. This means that a realistic model must be finite and saturated in the infrared region. In this work, we use the simplified form of Qin-Chang model \cite{Qin:2011dd}
\begin{align}
\mathcal{D}\left(k_{\Omega}^{2}\right)=\xi \frac{8 \pi^{2}}{\omega^{4}} e^{-k_{\Omega}^{2} / \omega^{2}}\,,
\end{align}
which have two parameters, i.e., the strength $\xi$ and the width $\omega$, for characterizing the interaction. Herein, we choose $\xi\omega=(0.82 \mathrm{GeV})^{3}$ with $\omega=0.5$ GeV. Since the simplified model has no ultraviolet tail, the renormalization procedure can be skipped. 

The solution of gap equation, i.e., the dressed quark propagator, can be decomposed as
\begin{align}
S\left({\omega}_{n}, \vec{p}\right)^{-1}= i\vec{\gamma} \cdot \vec{p} A +i \gamma_{4} {\omega}_{n} C + B\,,
\end{align}
where $A,B,C$ are scalar functions depending on the Matsubara frequency $\omega_n^2$ and the spatial momentum $\vec{p}^2$. With the solution for the fully dressed-quark propagator, we are able to study the chiral phase transition and obtain the (pseudo-)critical temperature $T_c$. The corresponding order parameter is the chiral condensate. In the chiral limit, we have
\begin{align}
-\langle\bar{\psi} \psi\rangle_{0} = N_{c}T \operatorname{tr}_{D}\sumint_n \frac{d^{3} p}{(2 \pi)^{3}} S\left(\vec{p}, \tilde{\omega}_{n}\right)
\label{qc}
\end{align}
Herein, the integral simply converges because the interaction model is ultraviolet-free. However, if a nonzero current quark mass is involved, Eq. \eqref{qc} diverges and does not give a well-defined chiral condensate. Hence, we introduce a subtraction procedure and define the chiral condensate as
\begin{align}
-\langle\bar{\psi} \psi\rangle=N_{c}T\operatorname{tr}_{D} \sumint_n\frac{d^{3} p}{(2 \pi)^{3}}\left[S\left(\vec{p}, \tilde{\omega}_{n}\right)-S_{0}\left(\vec{p}, \tilde{\omega}_{n}\right)\right]
\end{align}
with $S_{0}\left(\vec{p}, \tilde{\omega}_{n}\right)$ being the free quark propagator. It should be emphasized that the subtraction procedure only works for the simplified interaction model. More general discussions are available in Refs.  \cite{Williams:2006vva,Fischer:2009wc,Blank:2010bz,Fischer:2014ata,Contant:2017gtz,Gao:2016qkh}.

In this work, we choose the current quark masses
\begin{align}
m_{u/d} = 4.8\ \text{MeV} \text{ and } m_{s} = 116\ \text{MeV},
\end{align}
by fitting the pseudoscalar mesons in vacuum ($m_\pi = 0.135\ \text{GeV}$ and $m_K = 0.495\ \text{GeV}$). Accordingly, we obtain the leptonic decay constants ($f_{\pi} =0.095\ \text{GeV}$ and $f_{K} =0.110\ \text{GeV}$) which are well consistent with the empirical values. In the following discussion, we only analyze the chiral condensate of light quarks. For the case of strange quarks, see Refs. \cite{Fischer:2012vc,Fischer:2014ata,Fischer:2018sdj}, for instance.

\begin{figure}
	\centering
	\includegraphics[width=0.45\textwidth]{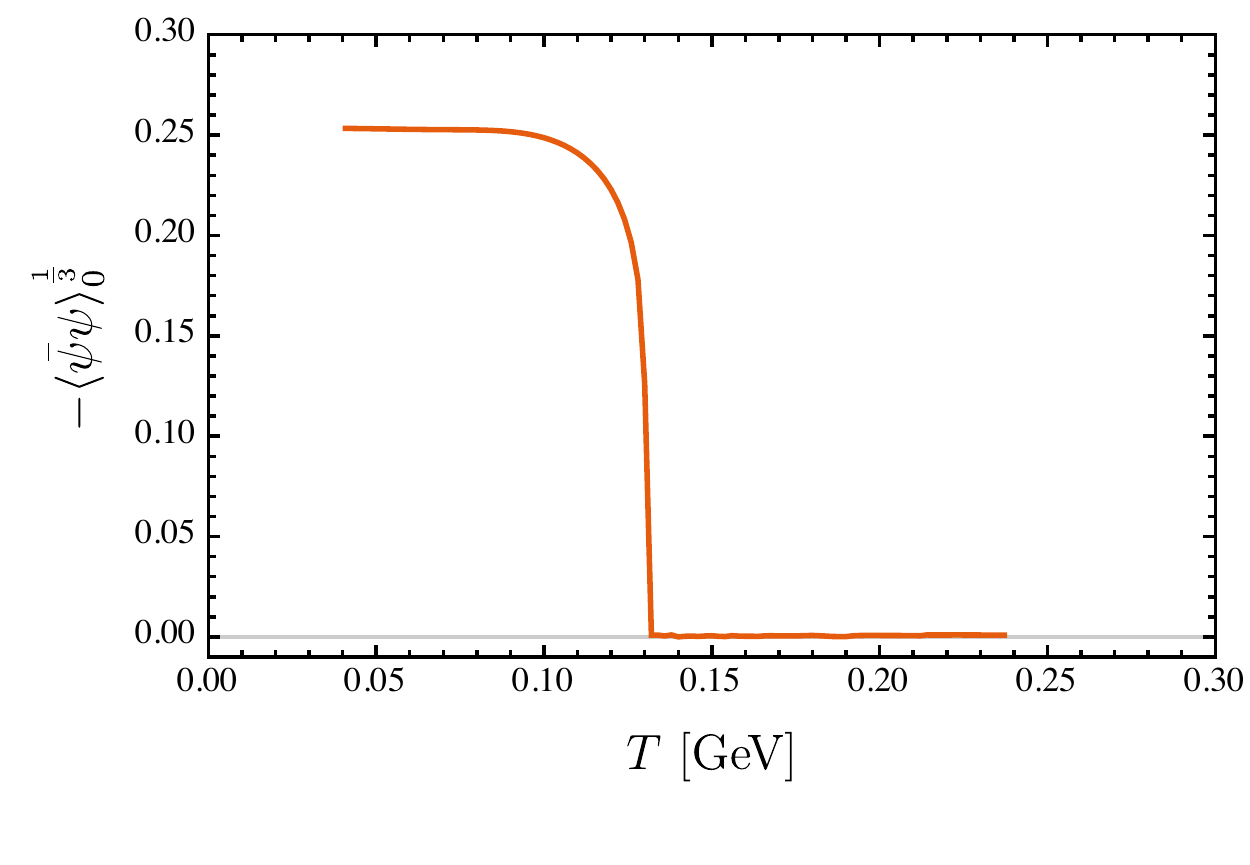}
	\includegraphics[width=0.45\textwidth]{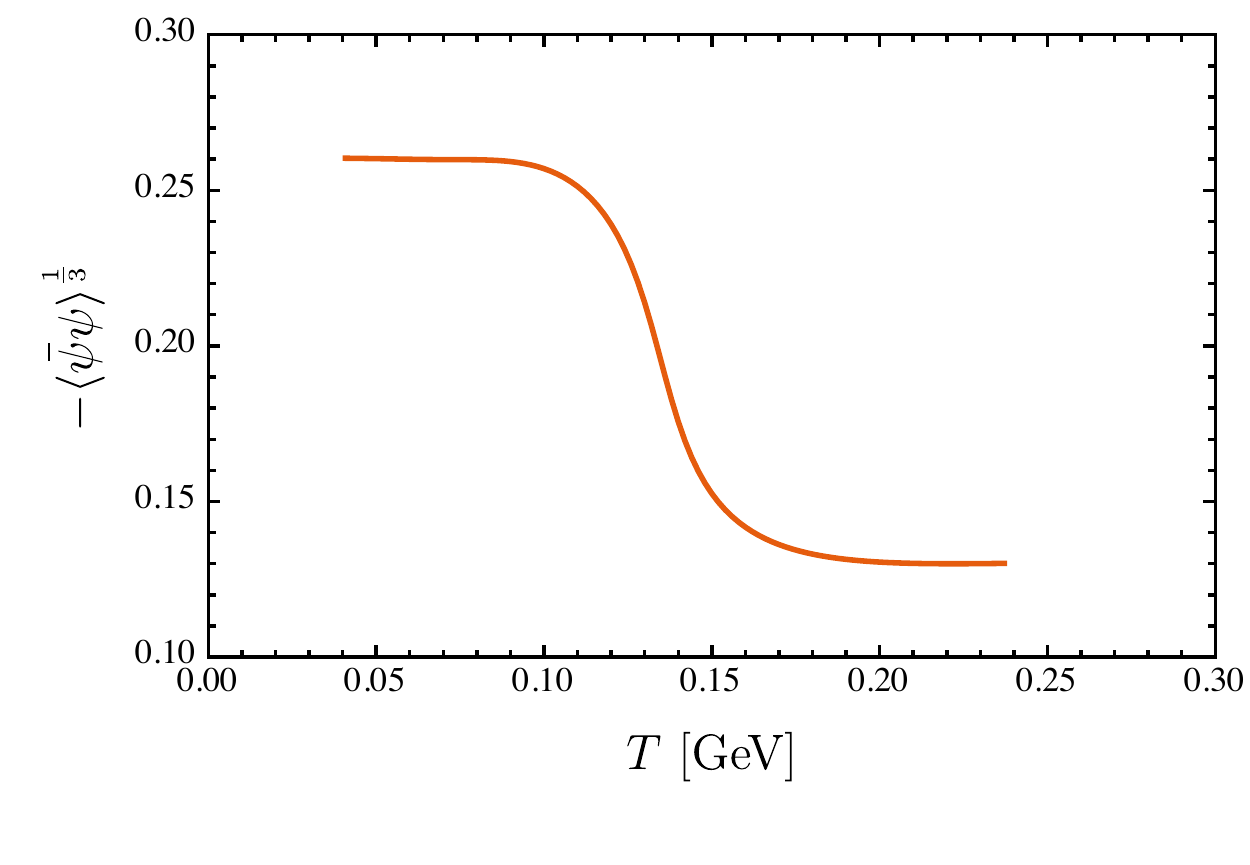}
	\vspace{-0.4cm}
	\caption{The thermal evolution behaviors of the chiral condensate in and beyond the chiral limit. Upper panel: $m_{u/d}=0$ MeV; Lower panel: $m_{u/d}=4.8$ MeV.}
	\label{figqc1}
\end{figure}

\begin{figure}
	\centering
	\includegraphics[width=0.45\textwidth]{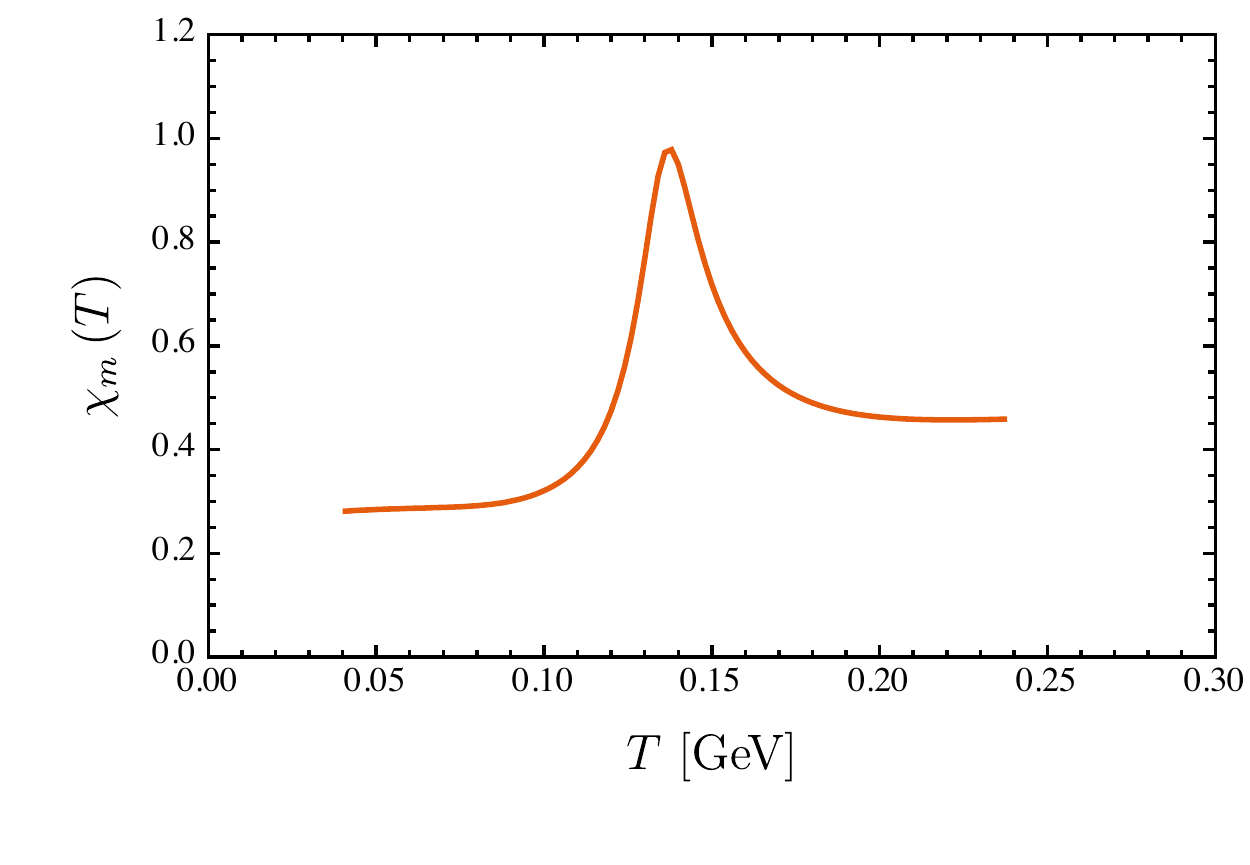}
	\vspace{-0.4cm}
	\caption{The thermal evolution behaviors of the chiral susceptibility beyond the chiral limit, where $m_{u/d} = 4.8$ MeV.}
	\label{figqc2}
\end{figure}

At zero temperature, we obtain the chiral condensate $-\langle\bar{\psi} \psi\rangle=(0.26\ \text{GeV})^3$, which is consistent with Gell-Mann-Oakes-Renner (GMOR) relation \cite{Maris:1997tm}:
\begin{align}
	f_\pi^2 m_\pi^2 = - 2 m_{u/d} \langle\bar{\psi} \psi\rangle
\end{align}
With temperature increasing, the chiral condensate in the chiral limit gradually decreases and completely vanishes when $T > 0.131$ MeV. The evolution behavior is shown in Fig. \ref{figqc1}. In other words, the chiral symmetry is partially restored at finite temperature. The complete restoration of chiral symmetry is a phase transition with the critical temperature
\begin{align}
 T_c\sim131\ \text{MeV}\,,
\end{align}
which is well consistent with the result in the chiral limit extrapolated by the latest lattice simulations: $T_{c}=132_{-6}^{+3}\ \text{MeV}$ \cite{HotQCD:2019xnw}. However, beyond the chiral limit, e.g., $m_{u/d}=4.8$ MeV, the phase transition of chiral symmetry restoration becomes a crossover (see the lower panel in Fig. \ref{figqc1}). For such case, the definition of the pseudo-critical temperature, i.e., $T_{pc}$, is not unique. For instance, $T_{pc}$ can be the temperature of the maximum chiral susceptibility \cite{Xu:2020loz,Fischer:2012vc,Fischer:2014ata} or the inflection point of the chiral condensate \cite{Steinbrecher:2018phh,Borsanyi:2012rr}. By studying the chiral susceptibility maximum, which is shown in Fig. \ref{figqc2}, we can obtain the pseudo-critical temperature
  \begin{align}
  T_{pc}\sim137\ \text{MeV}
  \end{align}
which is slightly lower than the recent result of lattice simulations with physical current quark mass: $T_{c}=156.5 \pm 1.5\ \text{MeV}$ \cite{HotQCD:2018pds}. 

\section{Bethe-Salpeter equation and meson properties}\label{sec:bse}
In the previous section, the quark gap equation is solved to study the chiral restoration at finite temperature, and the (pseudo-)critical temperatures are obtained. Now we investigate the properties of mesons, especially pion, in the neighborhood of $T_{c,pc}$. In vacuum, mesons can be solved by the homogeneous BSE, a counterpart of the Schr\"{o}dinger's equation. However, as mentioned in the introduction, it may be ill-defined at finite temperature. In this work, we follow Refs. \cite{Qin:2013aaa,Qin:2014dqa,Chen:2020afc} to extract meson properties from correlation functions.

In terms of Green functions, the Euclidean mesonic correlation function can be defined as \cite{Qin:2013aaa}
\begin{align}
\Pi_H(P_4, \vec{P}) \quad =\quad \includegraphics[valign=c,height=0.6cm,width=2.5cm]{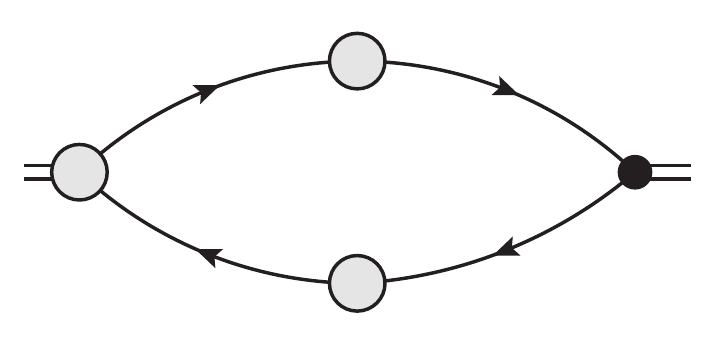}\ ,\label{cor}
\end{align}
where $P_4=2n\pi T$ with $n \in \mathbb{Z}$ is the bosonic Matsubara frequencies, and $\vec{P}$ is the total spatial momentum; gray circular blobs and black dots denote dressed vertices $\Gamma_H$ and bare vertices $\gamma_H$, respectively; the subscript $H$ specifies $J^P$ quantum number with $\gamma_H \in \{ \mathbf{I}, \gamma_5, \gamma_\mu, \gamma_5\gamma_\mu, \cdots \}$. The dressed vertices can be solved by the inhomogeneous BSE which reads
\begin{align}
\includegraphics[valign=c,height=0.6cm,width=5cm]{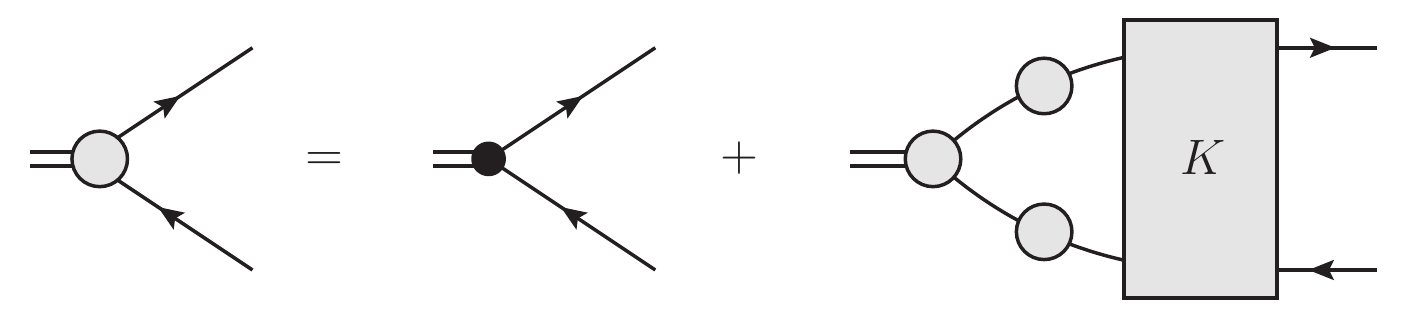} \,,
\label{BS}
\end{align}
where the dressed quark propagators are fed with the solution of the quark gap equation, and $K$ stands for the two-particle irreducible quark-antiquark scattering kernel.

In the rainbow-ladder approximation, the inhomogeneous BSE for the dressed vertices can be written as 
\begin{align} 
	&\Gamma_{H}\left(P_4; \tilde{\omega}_n, \vec{p}\right)=Z_{H} \gamma_{H}-Z_{1} \sumint_l \frac{d^{3} \vec{q}}{(2 \pi)^{3}} g^{2} D_{\mu \nu}\left(k_\Omega\right)\nonumber\\
	&~\times \frac{\lambda_{a}}{2} \gamma_{\mu} S\left(\tilde{\omega}_l, \vec{q}\right) \Gamma_{H}\left(P_4;\tilde{\omega}_l, \vec{q}\right) S\left(\tilde{\omega}_l+P_4, \vec{q}\right) \frac{\lambda_{a}}{2} \gamma_{v}\,, \label{bse}
\end{align}
where $\tilde{\omega}_n$ and $\vec{p}$ are the relative Matsubara frequencies and spatial momenta between two quarks, respectively, and the total spatial momenta of the vertices $\vec{P}=0$ since we can utilize the rest frame to extract observables without loss of generality. Since all momenta are Euclidean and real, we do not need any information of quark propagators on the complex plane. With the solutions of the quark gap equation and the inhomogeneous BSE, all ingredients in Eq. (\ref{cor}) are available, and thus we are able to compute the correlation functions. 

Observables are encoded in the spectral functions of correlation functions. The key to obtain observables is twofold: a well-defined spectral representation to bridge the spectral functions and the correlation functions; a reliable method to extract the spectral functions from the representation. First, in the rest frame, the correlation functions only depend on the Matsubara frequencies $P_4$ which are discrete. With polynomial interpolations of quark propagators, we are able to solve the vertices for $P_4 \in \mathbb{R}$ by using the inhomogeneous BSE, i.e., Eq. \eqref{bse}. Inserting the solution into Eq. \eqref{cor}, we then obtain the correlation functions on the whole real axis. We test several interpolation schemes for quark propagators and find that the produced correlation functions remain almost unchanged. In other words, the procedure is very accurate and robust, which, as we will see, could make our further analysis more convenient.

The correlation functions have a formal integral representation as
\begin{align}\Pi_{H}\left(P_4^{2}\right)=\int_{0}^{\infty} \frac{d \omega^{2}}{2 \pi} \frac{\rho_{H}(\omega)}{\omega^{2}+P_4^{2}} - (\text {subtraction})\,,
\label{rho}
\end{align}
where $\rho_H(\omega)$ is the spectral function. For the pseudoscalar (scalar) channel, in the non-interacting limit, the spectral function has an asymptotic behavior at large $\omega$ as \cite{Karsch:2003wy,Aarts:2005hg}
\begin{align}
\rho(\omega)\rightarrow \frac{3}{4\pi}\omega^2\label{asy}\,.
\end{align}
Since the spectral integral generates a quadratic divergence, an appropriate subtraction is required. 

Following Ref. \cite{Qin:2013aaa}, we introduce a discrete transform for correlation functions as (the subscript $H$ has been suppressed)
\begin{align}
	\hat{\Pi}\left(P_4^{2}\right) := & \frac{\Pi(s^{2})}{(s^{2}-s_1^{2})(s^{2}-s_2^{2})}\notag\\
&+\frac{\Pi(s_1^{2})}{(s_1^{2}-s^{2})(s_1^{2}-s_2^{2})}\notag\\
&+\frac{\Pi(s_2^{2})}{(s_2^{2}-s^{2})(s_2^{2}-s_1^{2})}\,.
\end{align}
where $s=P_4, s_1=P_4+\Delta, s_2=P_4+2\Delta$, and $\Delta$ is a positive constant. Then, the transformed correlation functions have the well-defined spectral representation as
\begin{align}
\hat{\Pi}\left(P_4^{2}\right)=\int_{0}^{\infty} \frac{d \omega^{2}}{2 \pi} \frac{\rho(\omega)}{\left(\omega^{2}+s^{2}\right)\left(\omega^{2}+s_1^2\right)\left(\omega^{2}+s_2^{2}\right)},\label{transform}
\end{align}
In principle, the choice of the constant $\Delta$ does not affect the extracted spectral functions. But in the numerical calculations, a large $\Delta$ could produce $\hat{\Pi}\left(P_4^{2}\right)$ with a small magnitude, which may involve sizable relative errors. Meanwhile, for a small $\Delta$, the transform involves small denominators and thus unstable numerical differences. In this work, we choose $\Delta = 0.8\pi \sim 4\pi T_{\rm max}$, where $T_{\rm max}=0.2$ GeV is the highest temperature in the analysis.

\begin{figure}
	\centering
	\includegraphics[width=0.45\textwidth]{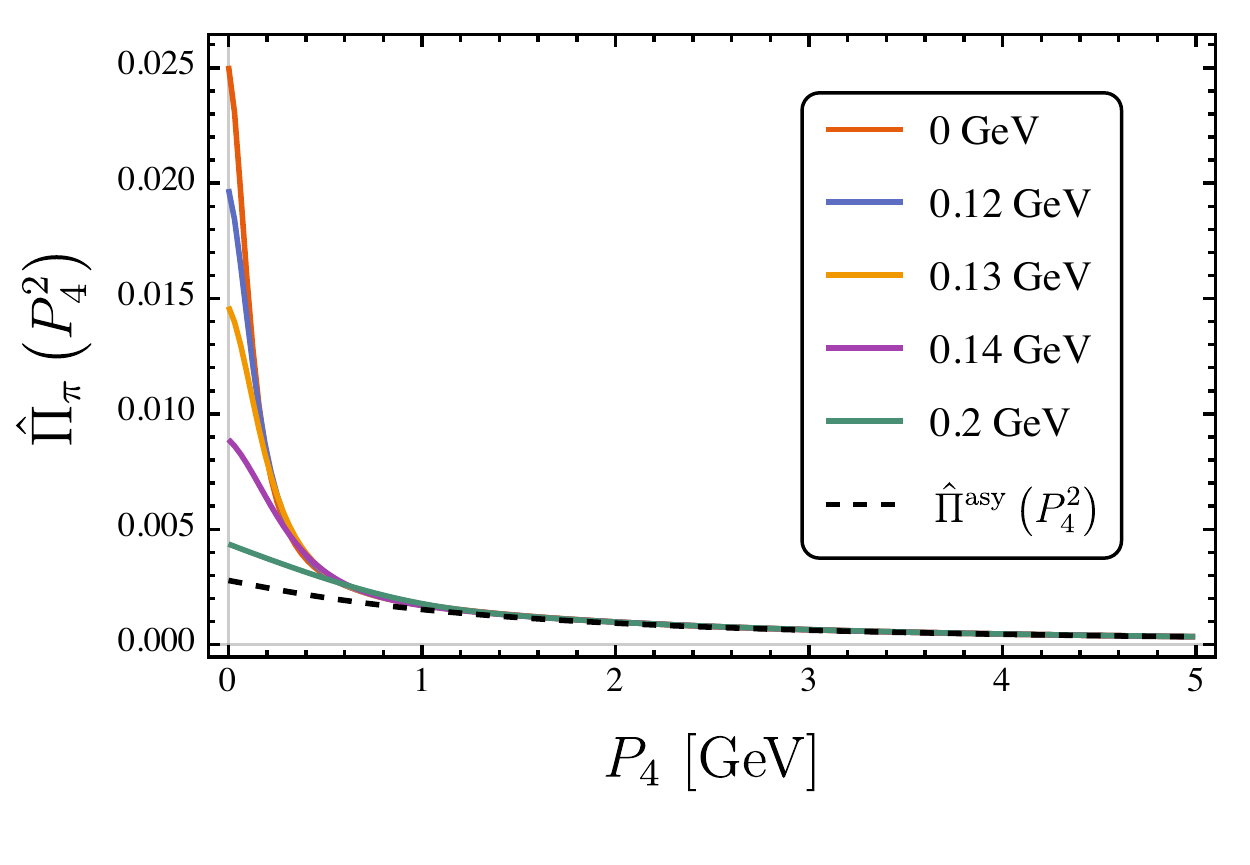}
	\vspace{-0.4cm}
	\caption{The correlation function of pion at different temperatures, where the current quark mass $m_{u/d}=4.8$ MeV.}
	\label{figcor}
\end{figure}

Inserting the non-interacting spectral function into the spectral representation Eq. \eqref{transform}, we then obtain the non-interacting correlation function as 
\begin{align}
\hat{\Pi}^{\text{asy}}\left(P_4^{2}\right)=\frac{3}{4\pi}\int_{0}^{\infty} \frac{d \omega^{2}}{2 \pi} \frac{\omega^2}{\left(\omega^{2}+s^{2}\right)\left(\omega^{2}+s_1^2\right)\left(\omega^{2}+s_2^{2}\right)}.\end{align}
By comparing the numerical results of $\hat{\Pi}_\pi(P_4^2)$ and $\hat{\Pi}^{\text{asy}}(P_4^2)$, we can intuitively see how the interaction is modified by heat bath. The comparison is presented in Fig. \ref{figcor}. It is shown that the transformed correlation functions $\hat{\Pi}_\pi$ are almost identical to that in vacuum when $T<0.1$ GeV. With temperature increasing, the infrared behaviors (i.e., $P_4<1$ GeV) of $\hat{\Pi}_\pi$ gradually decreases. This means that the low-energy interaction is weakened by heat bath. Moreover, it can be seen that the thermal effect is dramatically strong in the neighborhood of $T\sim 0.13$ GeV, which coincides the pseudo-critical temperature $T_{pc}$ of chiral phase crossover. Therefore, it is expected that the thermal evolution of meson properties, especially $\pi$ and $\sigma$, may strongly signal chiral phase transition or crossover.

\begin{figure}
	\centering
	\includegraphics[width=0.45\textwidth]{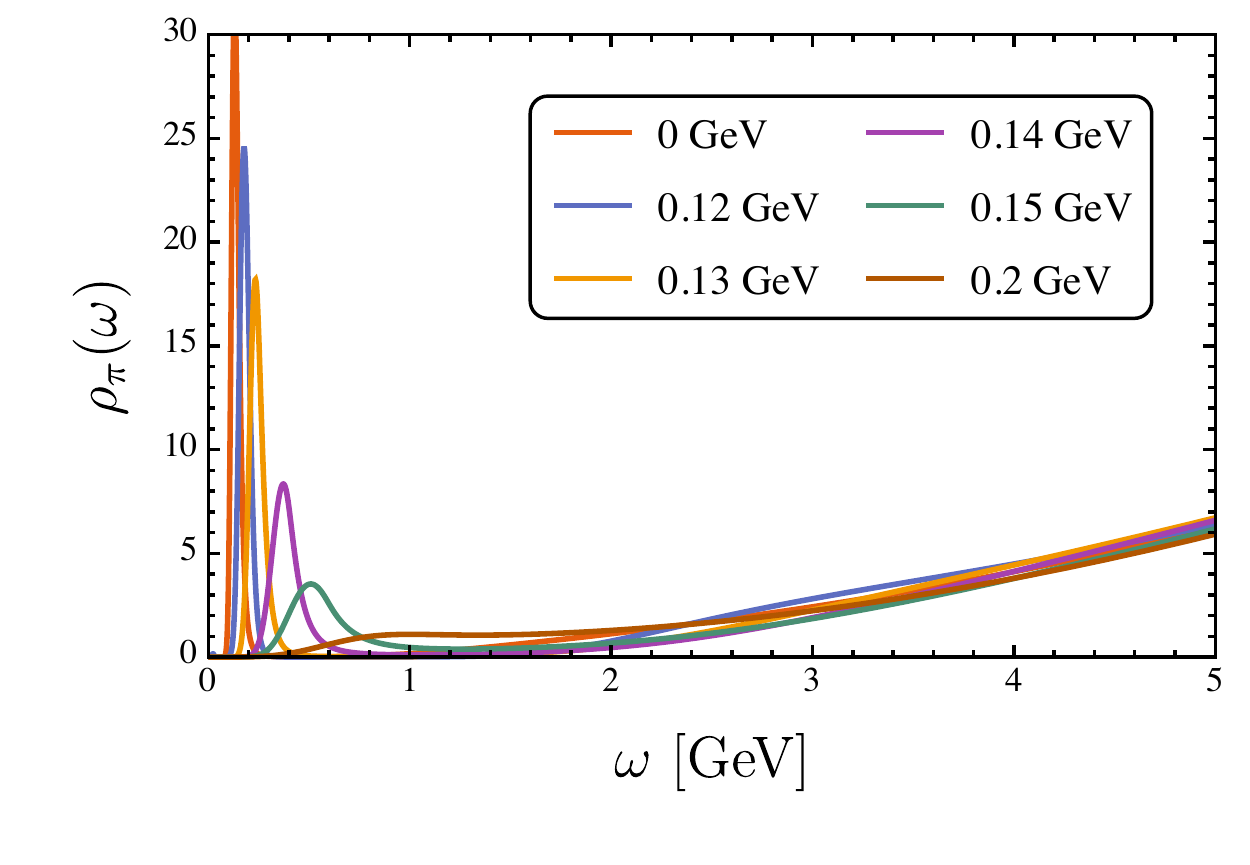}
	\vspace{-0.4cm}
	\caption{Extracted pion spectral functions at different temperatures, where the current quark mass $m_{u/d} = 4.8$ MeV.}
	\label{pirho}
\end{figure}

In order to obtain more information on mesons, such as mass and width, we still need to explicitly reconstruct spectral functions using Eq. \eqref{transform}. Although the reconstruction is an ill-posed problem, there are many numerical methods available on the market, e.g., maximum entropy method (MEM) that has been widely used \cite{Bryan:1990mx,Nickel:2006mm,Asakawa:2000tr}. In this work, we adopt MEM with the asymptotic spectral function as the default model. At the first place, it is found that a peak appears in the low-energy region of the extracted spectral functions, which is absent for the non-interacting spectral function. So the peak encodes information on the interaction. With temperature increasing,  the peaks become lower and boarder, and eventually disappear. For instance, the thermal evolution behavior for pion spectral functions is shown in Fig. \ref{pirho}.
 
In order to isolate the peak for analyzing meson properties, we can define the interacting spectral function as
\begin{align}
	\hat{\rho}(\omega) = {\rho}(\omega) - {\rho}^{\rm asy}(\omega)\,.
\end{align}
Then, the location of the peak corresponds to the meson mass, and the full width at half maximum (FWHM) to the decay width. The ratio of width to mass, denoted by $R$, can reflect how reliable a meson is defined. If the width is comparable to the mass, the peak can be hardly defined as a bound state. In the rest of this section, we analyze how the masses and the widths of the chiral partner, i.e., $\pi$ and $\sigma$ evolve with the temperature, and discuss how the thermal properties of Goldstone boson connect to chiral phase transition or crossover. 

\begin{figure}
	\centering
	\includegraphics[width=0.45\textwidth]{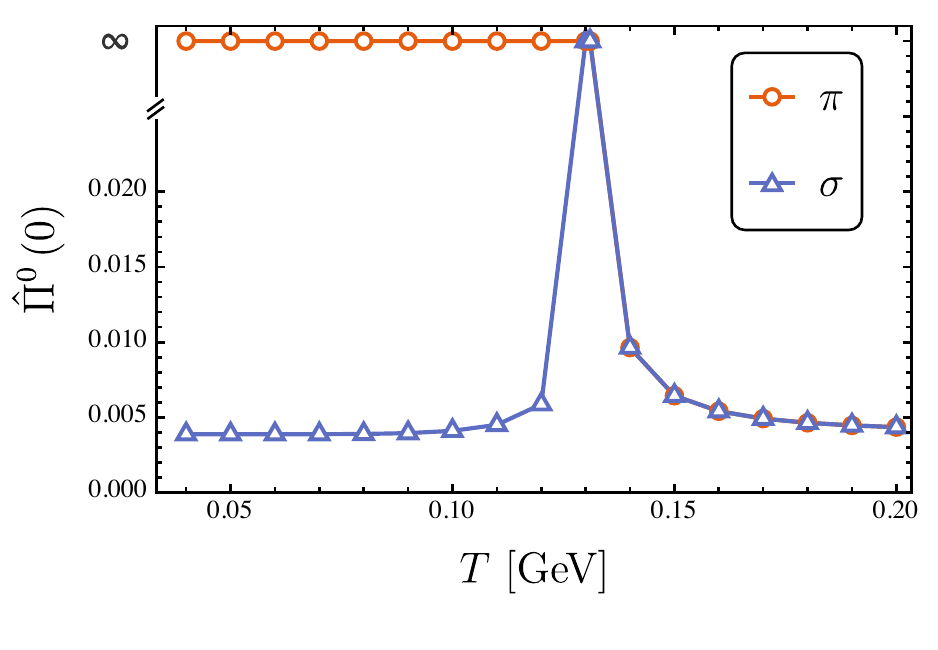}
	\includegraphics[width=0.45\textwidth]{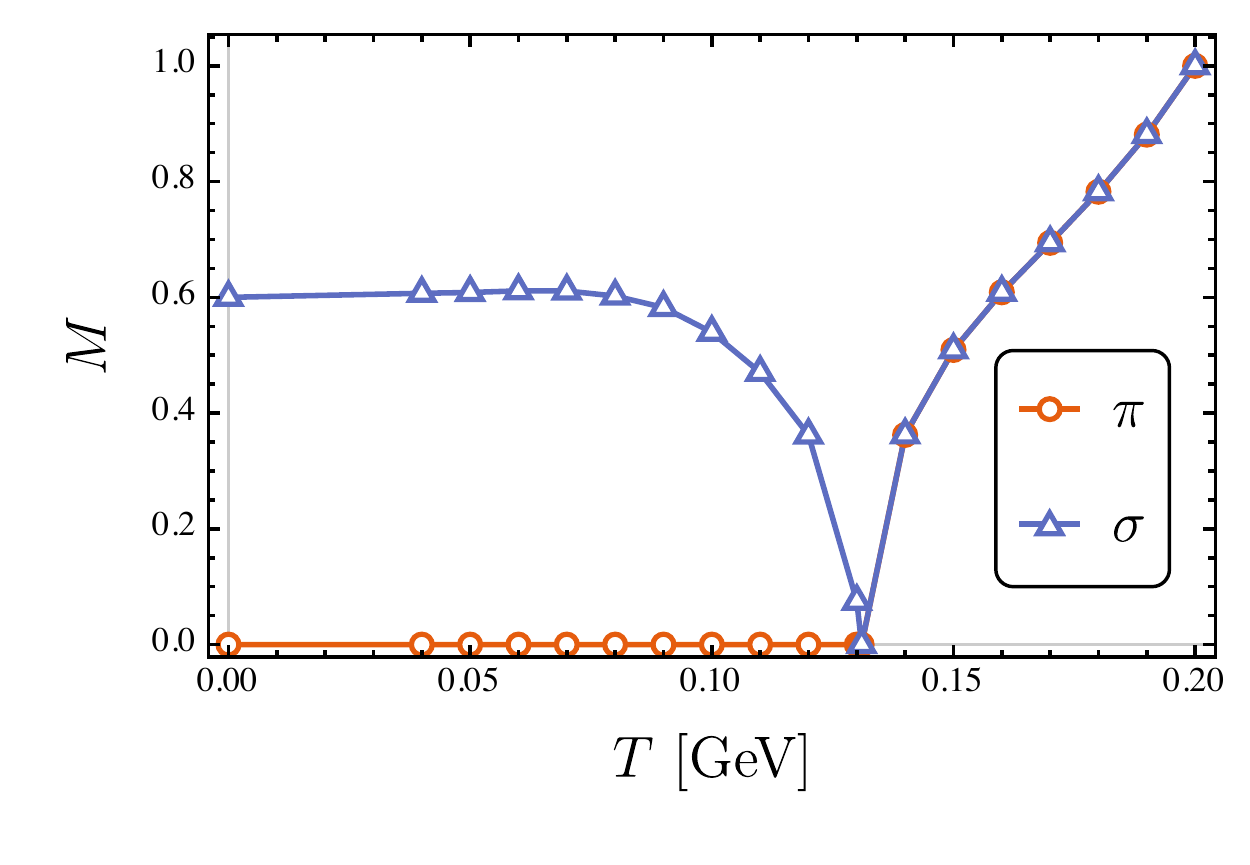}
	\includegraphics[width=0.45\textwidth]{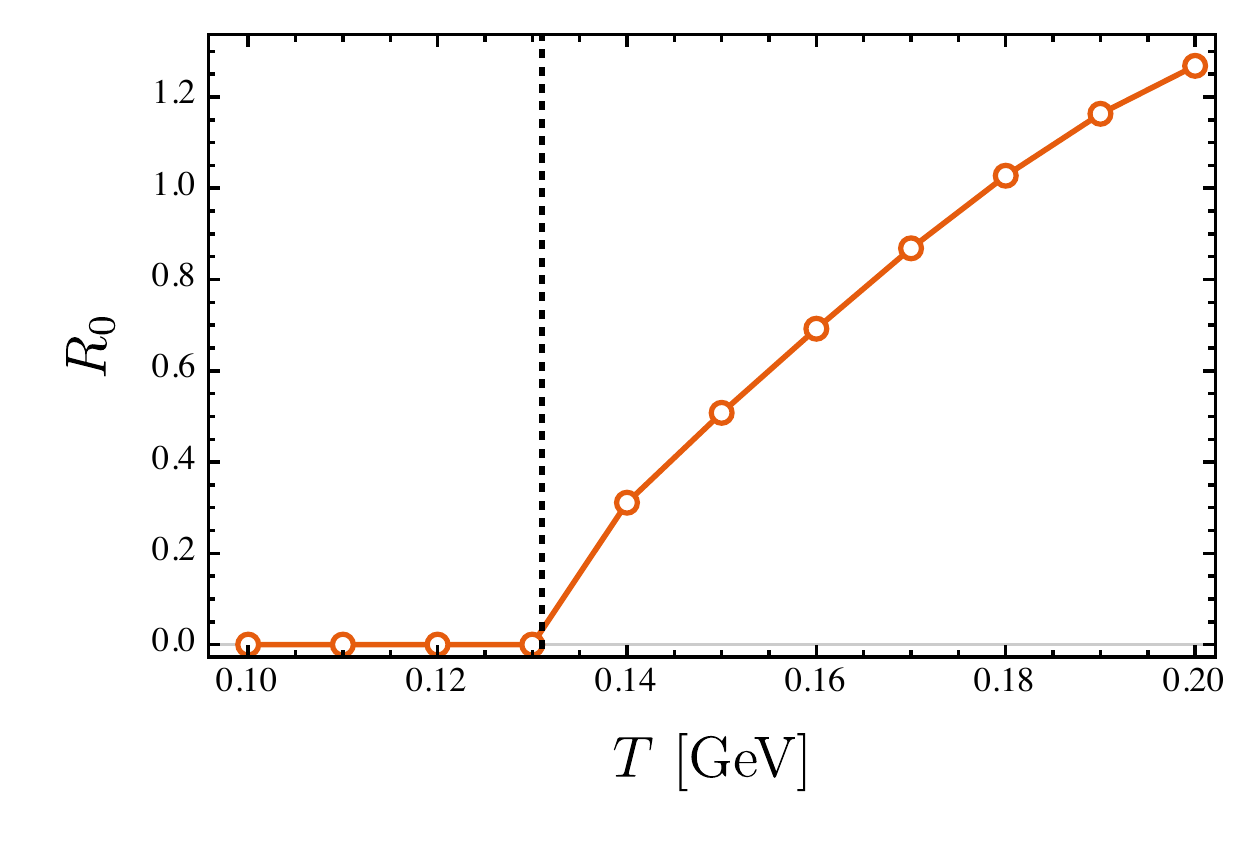}
	\vspace{-0.4cm}
	\caption{Upper panel: The pseudoscalar and scalar correlation functions at $P_4=0$ and different temperatures; Middle panel: The peak locations of pseudoscalar and scalar spectral functions at different temperatures; Lower panel: The ratios of width to mass for pseudoscalar peaks at different temperatures, where the vertical line indicates the steepest ascent temperature. Here, the calculation is performed in the chiral limit.}
	\label{figmass1}
\end{figure}

Let us first consider the case in the chiral limit. On the basis of the Goldstone theorem at finite temperature \cite{Yagi:2005yb}, the pion spectral function $\rho_\pi(\omega)$ should have a zero mass peak with vanishing width, which corresponds to massless pion, i.e.,
\begin{align}
\rho_\pi(\omega) \propto \delta(\omega) + {\rm regular~term}\,.
\end{align}
Accordingly, the pseudoscalar correlation function has a pole at $P_4 = 0$. The massless pion remains unchanged below the critical temperature, i.e., $T<T_c$. However, since the chiral symmetry is completely restored above the critical temperature, i.e., $T>T_c$, the zero mass peak for pion should shift. Moreover, properties of the chiral partner, i.e., $\pi$ and $\sigma$, must be identical due to the chiral symmetry. The numerical results are presented in Fig. \ref{figmass1} and confirm the aforementioned analysis. At $T>T_c$, the pion mass exhibits an increasing trend. It seems that pion can survive far above the critical temperature. However, it is found that the ratio of width to mass increases very rapidly after the chiral phase transition. This means that the pion cannot be considered as a well-defined bound state any more. We analyze the evolution of the ratio against temperature and find that the steepest ascent point $T_d$, which defines the pion dissociation temperature, coincides with $T_c$. The result is shown as the lower panel of Fig. \ref{figmass1}. Moreover, the pion dissociation can signal the deconfinement phase transition since quarks and antiquarks can be released from bound states. In other words, it can be concluded that the chiral and deconfinement phase transitions happen at the same temperature. The conclusion is consistent with the analysis of quark spectral properties \cite{Mueller:2010ah,Qin:2013ufa}.

\begin{figure}
	\centering
	\includegraphics[width=0.45\textwidth]{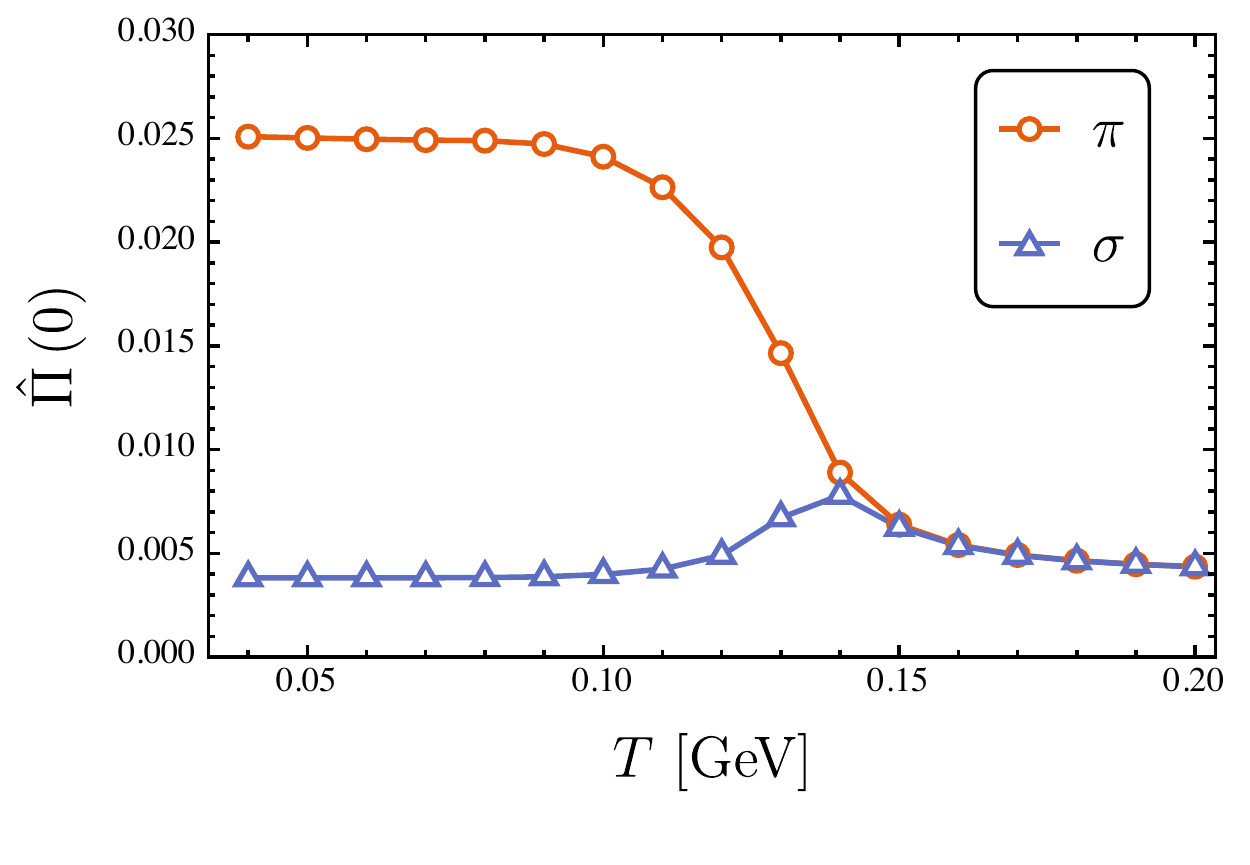}
	\includegraphics[width=0.45\textwidth]{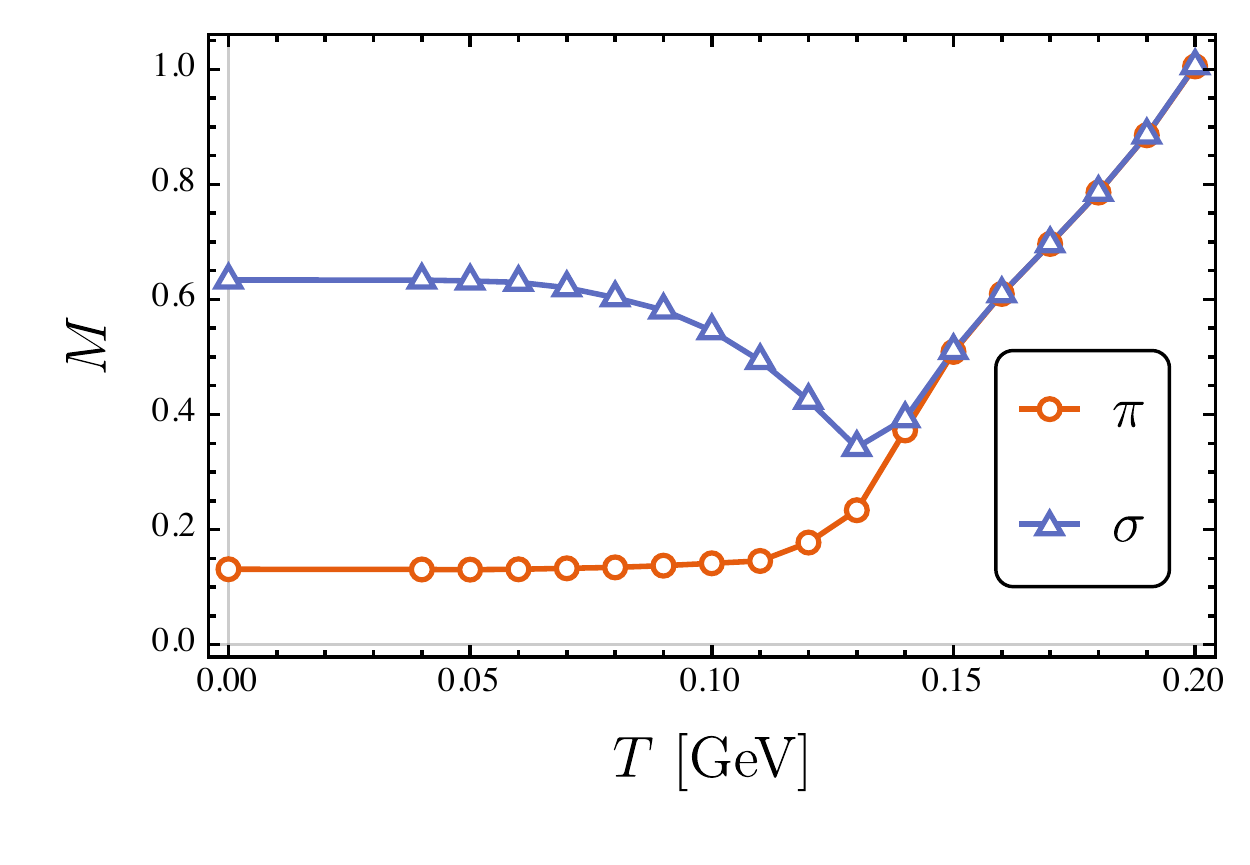}
	\includegraphics[width=0.45\textwidth]{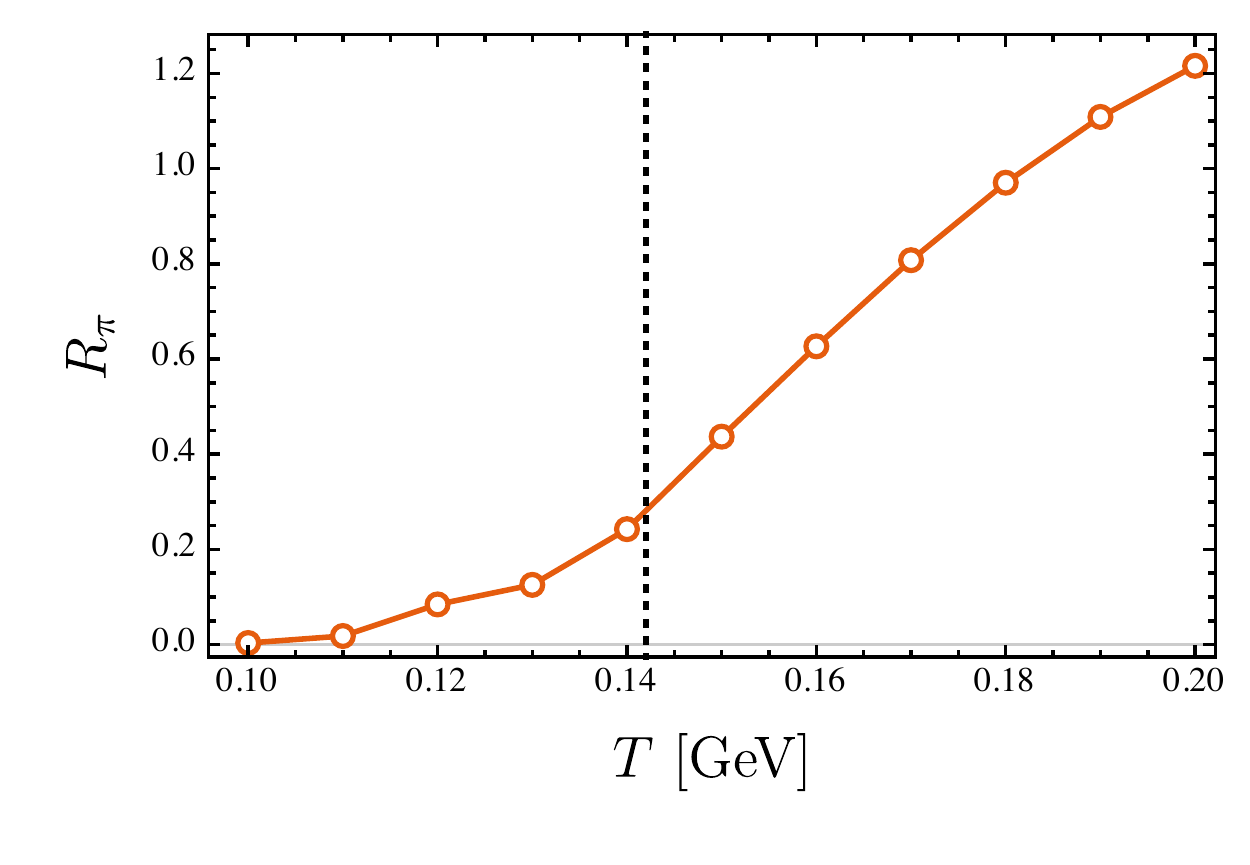}
	\vspace{-0.4cm}
	\caption{Upper panel: The pseudoscalar and scalar correlation functions at $P_4=0$ and different temperatures; Middle panel: The peak locations of pseudoscalar and scalar spectral functions at different temperatures; Lower panel: The ratios of width to mass for pseudoscalar peaks at different temperatures, where the vertical line indicates the steepest ascent temperature. Here, the calculation is performed beyond the chiral limit, i.e., $m_{u/d} = 4.8$ MeV.}
	\label{figmass2}
\end{figure}

Now we discuss the case beyond the chiral limit, i.e., $m_{u/d} = 4.8$ MeV. It is generally believed that, once a small current quark mass is involved, the chiral symmetry is no longer exact, and the pion becomes slightly massive according to the GMOR relation. It is expected that the pion has a finite width at finite temperature. The disappearing of massless pion can be easily read out from the divergence-free correlation functions (see for instance the upper panel of Fig. \ref{figmass2}). With temperature increasing, the difference between pseudoscalar and scalar correlation functions decreases, but do not eventually vanish in the asymptotic limit since the chiral symmetry is explicitly broken by the finite current quark mass. For $T\lesssim m_\pi$, the masses of $\pi$ and $\sigma$ increases and decreases, respectively. They become very close at high temperature, e.g., $T>0.15$ GeV. The evolution behaviors of the $\pi$ and $\sigma$ masses against temperature is shown as the middle panel of Fig. \ref{figmass2}. Our result is consistent with that obtained by the ChPT which points out a slight positive shift of pion mass at low temperature \cite{Yagi:2005yb,Toublan:1997rr}. 

As mentioned above, for the chiral phase crossover, the definition of pseudo-critical temperature is not unique. Based on the discussion in the chiral limit, we again study the pion dissociation by analyzing its ratio of width to mass against temperature. Then, we can define the pseudo-critical temperature as the pion dissociation temperature, i.e.,
\begin{align}
	T_{pc} := T_d^\pi \,.
\end{align}
The result is shown as the lower panel of Fig. \ref{figmass2}, from which it is obtained that $T_{pc} = T_{d}^\pi = 0.142$ GeV. This value is slightly higher than that defined by the maximum point of the chiral susceptibility. Considering that numerical uncertainties may exist in the reconstruction of spectral functions, the deviation is negligible. In other words, the pion dissociation can signal phase transition, effectively.

\begin{figure}
	\centering
	\includegraphics[width=0.45\textwidth]{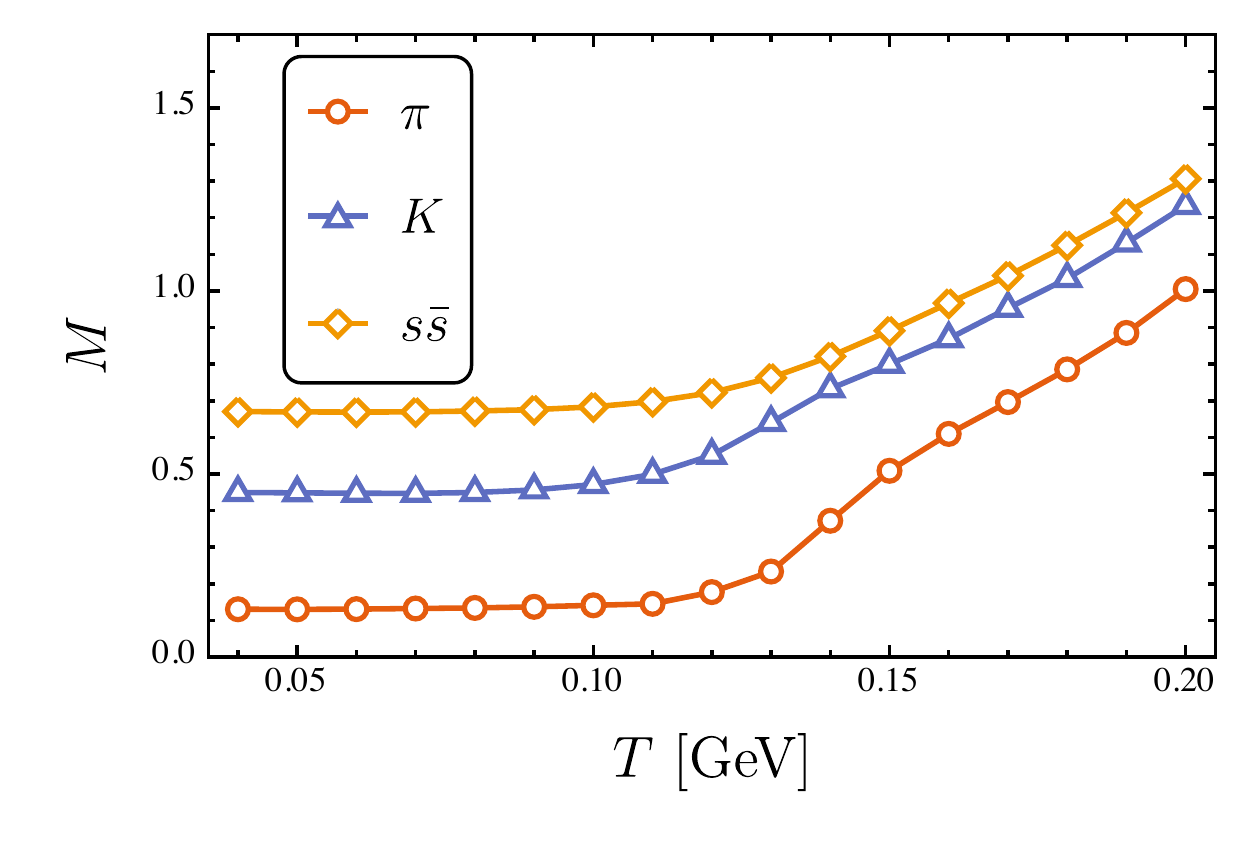}
	\includegraphics[width=0.45\textwidth]{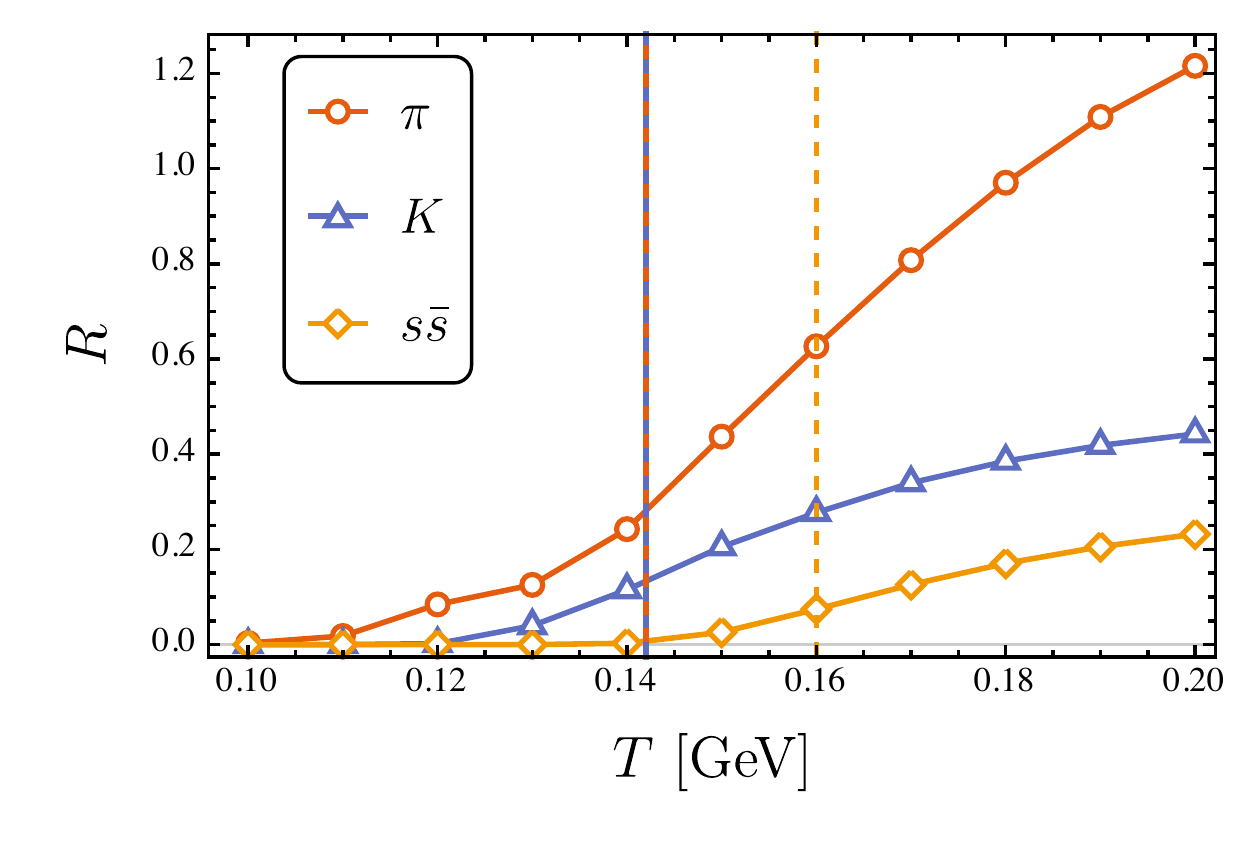}
	\vspace{-0.4cm}
	\caption{Upper panel: The mass evolutions of Goldstone modes with temperature; Lower panel: The ratios of width to mass for Goldstone modes at different temperatures, where the vertical line indicates the steepest ascent temperature.}
	\label{figmeson}
\end{figure}

In order to complete our discussion, we parallelly extend our analysis to the three-flavor case, which involves $u$, $d$ and $s$-quarks. Accordingly, thermal properties of the Goldstone modes, e.g., pion, kaon, and fictitious pseudoscalar $s\bar{s}$, are discussed. At low temperature, all modes remain almost untouched. With temperature increasing, their masses gradually increase, but the increasing behaviors are smoother for heavier modes. The comparison for all modes is shown as the upper panel of Fig. \ref{figmeson}. The lower panel of the same figure shows the ratios of width to mass for all modes at different temperatures. The vertical lines indicate the dissociation temperatures, i.e., $T_d^K \approx T_d^\pi = 0.142$ GeV and $T_d^{s\bar{s}} \approx 1.14 T_d^\pi = 0.160$ GeV. This suggests that bound states of heavy quarks may survive after the chiral restoration crossover (for similar results, see, e.g., Ref. \cite{Chen:2020afc} and references therein).

\section{Summary and outlook}\label{sec:summary}
In summary, we studied QCD phase transitions, i.e., chiral symmetry restoration and deconfinement, at finite temperature and zero chemical potential. We mainly focused on the thermal properties of pseudoscalar mesons, e.g., $\pi$, $K$, and pseudoscalar $s\bar{s}$, due to their twofold roles as bound states and Goldstone modes. The meson properties were extracted by reconstructing spectral functions from Euclidean mesonic correlation functions which can be computed with the self-consistent solutions of the rainbow-ladder truncated BSE and DSE. 

In the chiral limit, with temperature increasing, the chiral symmetry is restored via a phase transition at $T_{c} \sim 0.131$ GeV. For $T<T_c$, the pion is always massless as the Goldstone boson of DCSB. The comparison between the evolution behaviors of pseudoscalar and scalar correlation functions against temperature indicates the gradual chiral restoration. Moreover, the pion dissociates at the critical temperature. The pion dissociation temperature, denoted by $T_d^\pi$, can be considered as the critical temperature of both chiral symmetry restoration and deconfinement. Beyond the chiral limit, the pseudo-critical temperature is obtained by the maximum point of the chiral susceptibility, i.e., $T_{pc} = 0.137$ MeV. Considering that the definition of $T_{pc}$ is not unique, it is suggested that $T_d^\pi$ can serve as alternative option. For the physical current quark mass, $T_d^\pi = 0.142$ GeV, which is consistent with that obtained by analyzing the chiral susceptibility. At last, we analyzed thermal properties of the Goldstone modes for the three-flavor case. It is found that $\pi$ and $K$ dissociate almost simultaneously, but heavier $s\bar{s}$ can survive at higher temperature.

On the one hand, the above studies can be improved by using a sophisticated truncation scheme beyond the RL approximation \cite{Qin:2020jig}. On the other hand, the above analysis can be extended to the QCD phase diagram on the temperature and chemical potential plane. In-medium meson properties can serves as a powerful tool to identify phase structures and reveal phase properties.

\section*{Acknowledgements}
To memorize Hong-Shi Zong, a great mentor and close friend who passed away in March 2021. We are grateful for helpful discussion with Heng-Tong Ding, Hai-Tao Shu and Shi-Yang Chen. This work was supported in part by the National Natural Science Foundation of China (under Grants Nos. 12075117, 11775112, 11535005, 11690030, 11905104, 11805024, and 11947406) and the Jiangsu Provincial Natural Science Foundation of China under Grant No. BK20180323.


\begin{thebibliography}{57}%
\makeatletter
\providecommand \@ifxundefined [1]{%
 \@ifx{#1\undefined}
}%
\providecommand \@ifnum [1]{%
 \ifnum #1\expandafter \@firstoftwo
 \else \expandafter \@secondoftwo
 \fi
}%
\providecommand \@ifx [1]{%
 \ifx #1\expandafter \@firstoftwo
 \else \expandafter \@secondoftwo
 \fi
}%
\providecommand \natexlab [1]{#1}%
\providecommand \enquote  [1]{``#1''}%
\providecommand \bibnamefont  [1]{#1}%
\providecommand \bibfnamefont [1]{#1}%
\providecommand \citenamefont [1]{#1}%
\providecommand \href@noop [0]{\@secondoftwo}%
\providecommand \href [0]{\begingroup \@sanitize@url \@href}%
\providecommand \@href[1]{\@@startlink{#1}\@@href}%
\providecommand \@@href[1]{\endgroup#1\@@endlink}%
\providecommand \@sanitize@url [0]{\catcode `\\12\catcode `\$12\catcode
  `\&12\catcode `\#12\catcode `\^12\catcode `\_12\catcode `\%12\relax}%
\providecommand \@@startlink[1]{}%
\providecommand \@@endlink[0]{}%
\providecommand \url  [0]{\begingroup\@sanitize@url \@url }%
\providecommand \@url [1]{\endgroup\@href {#1}{\urlprefix }}%
\providecommand \urlprefix  [0]{URL }%
\providecommand \Eprint [0]{\href }%
\providecommand \doibase [0]{http://dx.doi.org/}%
\providecommand \selectlanguage [0]{\@gobble}%
\providecommand \bibinfo  [0]{\@secondoftwo}%
\providecommand \bibfield  [0]{\@secondoftwo}%
\providecommand \translation [1]{[#1]}%
\providecommand \BibitemOpen [0]{}%
\providecommand \bibitemStop [0]{}%
\providecommand \bibitemNoStop [0]{.\EOS\space}%
\providecommand \EOS [0]{\spacefactor3000\relax}%
\providecommand \BibitemShut  [1]{\csname bibitem#1\endcsname}%
\let\auto@bib@innerbib\@empty
\bibitem [{\citenamefont {Cabibbo}\ and\ \citenamefont
  {Parisi}(1975)}]{Cabibbo:1975ig}%
  \BibitemOpen
  \bibfield  {author} {\bibinfo {author} {\bibfnamefont {N.}~\bibnamefont
  {Cabibbo}}\ and\ \bibinfo {author} {\bibfnamefont {G.}~\bibnamefont
  {Parisi}},\ }\href {\doibase 10.1016/0370-2693(75)90158-6} {\bibfield
  {journal} {\bibinfo  {journal} {Phys. Lett. B}\ }\textbf {\bibinfo {volume}
  {59}},\ \bibinfo {pages} {67} (\bibinfo {year} {1975})}\BibitemShut {NoStop}%
\bibitem [{\citenamefont {Sarkar}\ \emph {et~al.}(2010)\citenamefont {Sarkar},
  \citenamefont {Satz},\ and\ \citenamefont {Sinha}}]{Sarkar:2010zza}%
  \BibitemOpen
  \bibinfo {editor} {\bibfnamefont {S.}~\bibnamefont {Sarkar}}, \bibinfo
  {editor} {\bibfnamefont {H.}~\bibnamefont {Satz}}, \ and\ \bibinfo {editor}
  {\bibfnamefont {B.}~\bibnamefont {Sinha}},\ eds.,\ \href {\doibase
  10.1007/978-3-642-02286-9} {\emph {\bibinfo {title} {{The physics of the
  quark-gluon plasma}}}},\ Vol.\ \bibinfo {volume} {785}\ (\bibinfo {year}
  {2010})\BibitemShut {NoStop}%
\bibitem [{\citenamefont {Fischer}(2019)}]{Fischer:2018sdj}%
  \BibitemOpen
  \bibfield  {author} {\bibinfo {author} {\bibfnamefont {C.~S.}\ \bibnamefont
  {Fischer}},\ }\href {\doibase 10.1016/j.ppnp.2019.01.002} {\bibfield
  {journal} {\bibinfo  {journal} {Prog. Part. Nucl. Phys.}\ }\textbf {\bibinfo
  {volume} {105}},\ \bibinfo {pages} {1} (\bibinfo {year} {2019})},\ \Eprint
  {http://arxiv.org/abs/1810.12938} {arXiv:1810.12938 [hep-ph]} \BibitemShut
  {NoStop}%
\bibitem [{\citenamefont {Fukushima}\ and\ \citenamefont
  {Hatsuda}(2011)}]{Fukushima:2010bq}%
  \BibitemOpen
  \bibfield  {author} {\bibinfo {author} {\bibfnamefont {K.}~\bibnamefont
  {Fukushima}}\ and\ \bibinfo {author} {\bibfnamefont {T.}~\bibnamefont
  {Hatsuda}},\ }\href {\doibase 10.1088/0034-4885/74/1/014001} {\bibfield
  {journal} {\bibinfo  {journal} {Rept. Prog. Phys.}\ }\textbf {\bibinfo
  {volume} {74}},\ \bibinfo {pages} {014001} (\bibinfo {year} {2011})},\
  \Eprint {http://arxiv.org/abs/1005.4814} {arXiv:1005.4814 [hep-ph]}
  \BibitemShut {NoStop}%
\bibitem [{\citenamefont {Roberts}\ and\ \citenamefont
  {Schmidt}(2000)}]{Roberts:2000aa}%
  \BibitemOpen
  \bibfield  {author} {\bibinfo {author} {\bibfnamefont {C.~D.}\ \bibnamefont
  {Roberts}}\ and\ \bibinfo {author} {\bibfnamefont {S.~M.}\ \bibnamefont
  {Schmidt}},\ }\href {\doibase 10.1016/S0146-6410(00)90011-5} {\bibfield
  {journal} {\bibinfo  {journal} {Prog. Part. Nucl. Phys.}\ }\textbf {\bibinfo
  {volume} {45}},\ \bibinfo {pages} {S1} (\bibinfo {year} {2000})},\ \Eprint
  {http://arxiv.org/abs/nucl-th/0005064} {arXiv:nucl-th/0005064} \BibitemShut
  {NoStop}%
\bibitem [{\citenamefont {Ding}\ \emph {et~al.}(2015)\citenamefont {Ding},
  \citenamefont {Karsch},\ and\ \citenamefont {Mukherjee}}]{Ding:2015ona}%
  \BibitemOpen
  \bibfield  {author} {\bibinfo {author} {\bibfnamefont {H.-T.}\ \bibnamefont
  {Ding}}, \bibinfo {author} {\bibfnamefont {F.}~\bibnamefont {Karsch}}, \ and\
  \bibinfo {author} {\bibfnamefont {S.}~\bibnamefont {Mukherjee}},\ }\href
  {\doibase 10.1142/S0218301315300076} {\bibfield  {journal} {\bibinfo
  {journal} {Int. J. Mod. Phys. E}\ }\textbf {\bibinfo {volume} {24}},\
  \bibinfo {pages} {1530007} (\bibinfo {year} {2015})},\ \Eprint
  {http://arxiv.org/abs/1504.05274} {arXiv:1504.05274 [hep-lat]} \BibitemShut
  {NoStop}%
\bibitem [{\citenamefont {Qin}\ \emph {et~al.}(2011{\natexlab{a}})\citenamefont
  {Qin}, \citenamefont {Chang}, \citenamefont {Chen}, \citenamefont {Liu},\
  and\ \citenamefont {Roberts}}]{Qin:2010nq}%
  \BibitemOpen
  \bibfield  {author} {\bibinfo {author} {\bibfnamefont {S.-x.}\ \bibnamefont
  {Qin}}, \bibinfo {author} {\bibfnamefont {L.}~\bibnamefont {Chang}}, \bibinfo
  {author} {\bibfnamefont {H.}~\bibnamefont {Chen}}, \bibinfo {author}
  {\bibfnamefont {Y.-x.}\ \bibnamefont {Liu}}, \ and\ \bibinfo {author}
  {\bibfnamefont {C.~D.}\ \bibnamefont {Roberts}},\ }\href {\doibase
  10.1103/PhysRevLett.106.172301} {\bibfield  {journal} {\bibinfo  {journal}
  {Phys. Rev. Lett.}\ }\textbf {\bibinfo {volume} {106}},\ \bibinfo {pages}
  {172301} (\bibinfo {year} {2011}{\natexlab{a}})},\ \Eprint
  {http://arxiv.org/abs/1011.2876} {arXiv:1011.2876 [nucl-th]} \BibitemShut
  {NoStop}%
\bibitem [{\citenamefont {Gao}\ and\ \citenamefont
  {Pawlowski}(2020)}]{Gao:2020qsj}%
  \BibitemOpen
  \bibfield  {author} {\bibinfo {author} {\bibfnamefont {F.}~\bibnamefont
  {Gao}}\ and\ \bibinfo {author} {\bibfnamefont {J.~M.}\ \bibnamefont
  {Pawlowski}},\ }\href {\doibase 10.1103/PhysRevD.102.034027} {\bibfield
  {journal} {\bibinfo  {journal} {Phys. Rev. D}\ }\textbf {\bibinfo {volume}
  {102}},\ \bibinfo {pages} {034027} (\bibinfo {year} {2020})},\ \Eprint
  {http://arxiv.org/abs/2002.07500} {arXiv:2002.07500 [hep-ph]} \BibitemShut
  {NoStop}%
\bibitem [{\citenamefont {Braun}\ \emph {et~al.}(2020)\citenamefont {Braun},
  \citenamefont {Fu}, \citenamefont {Pawlowski}, \citenamefont {Rennecke},
  \citenamefont {Rosenbl\"uh},\ and\ \citenamefont {Yin}}]{Braun:2020ada}%
  \BibitemOpen
  \bibfield  {author} {\bibinfo {author} {\bibfnamefont {J.}~\bibnamefont
  {Braun}}, \bibinfo {author} {\bibfnamefont {W.-j.}\ \bibnamefont {Fu}},
  \bibinfo {author} {\bibfnamefont {J.~M.}\ \bibnamefont {Pawlowski}}, \bibinfo
  {author} {\bibfnamefont {F.}~\bibnamefont {Rennecke}}, \bibinfo {author}
  {\bibfnamefont {D.}~\bibnamefont {Rosenbl\"uh}}, \ and\ \bibinfo {author}
  {\bibfnamefont {S.}~\bibnamefont {Yin}},\ }\href {\doibase
  10.1103/PhysRevD.102.056010} {\bibfield  {journal} {\bibinfo  {journal}
  {Phys. Rev. D}\ }\textbf {\bibinfo {volume} {102}},\ \bibinfo {pages}
  {056010} (\bibinfo {year} {2020})},\ \Eprint
  {http://arxiv.org/abs/2003.13112} {arXiv:2003.13112 [hep-ph]} \BibitemShut
  {NoStop}%
\bibitem [{\citenamefont {Fu}\ \emph {et~al.}(2020)\citenamefont {Fu},
  \citenamefont {Pawlowski},\ and\ \citenamefont {Rennecke}}]{Fu:2019hdw}%
  \BibitemOpen
  \bibfield  {author} {\bibinfo {author} {\bibfnamefont {W.-j.}\ \bibnamefont
  {Fu}}, \bibinfo {author} {\bibfnamefont {J.~M.}\ \bibnamefont {Pawlowski}}, \
  and\ \bibinfo {author} {\bibfnamefont {F.}~\bibnamefont {Rennecke}},\ }\href
  {\doibase 10.1103/PhysRevD.101.054032} {\bibfield  {journal} {\bibinfo
  {journal} {Phys. Rev. D}\ }\textbf {\bibinfo {volume} {101}},\ \bibinfo
  {pages} {054032} (\bibinfo {year} {2020})},\ \Eprint
  {http://arxiv.org/abs/1909.02991} {arXiv:1909.02991 [hep-ph]} \BibitemShut
  {NoStop}%
\bibitem [{\citenamefont {Brandt}\ \emph {et~al.}(2015)\citenamefont {Brandt},
  \citenamefont {Francis}, \citenamefont {Meyer},\ and\ \citenamefont
  {Robaina}}]{Brandt:2015sxa}%
  \BibitemOpen
  \bibfield  {author} {\bibinfo {author} {\bibfnamefont {B.~B.}\ \bibnamefont
  {Brandt}}, \bibinfo {author} {\bibfnamefont {A.}~\bibnamefont {Francis}},
  \bibinfo {author} {\bibfnamefont {H.~B.}\ \bibnamefont {Meyer}}, \ and\
  \bibinfo {author} {\bibfnamefont {D.}~\bibnamefont {Robaina}},\ }\href
  {\doibase 10.1103/PhysRevD.92.094510} {\bibfield  {journal} {\bibinfo
  {journal} {Phys. Rev. D}\ }\textbf {\bibinfo {volume} {92}},\ \bibinfo
  {pages} {094510} (\bibinfo {year} {2015})},\ \Eprint
  {http://arxiv.org/abs/1506.05732} {arXiv:1506.05732 [hep-lat]} \BibitemShut
  {NoStop}%
\bibitem [{\citenamefont {Cheng}\ \emph {et~al.}(2011)\citenamefont {Cheng}
  \emph {et~al.}}]{Cheng:2010fe}%
  \BibitemOpen
  \bibfield  {author} {\bibinfo {author} {\bibfnamefont {M.}~\bibnamefont
  {Cheng}} \emph {et~al.},\ }\href {\doibase 10.1140/epjc/s10052-011-1564-y}
  {\bibfield  {journal} {\bibinfo  {journal} {Eur. Phys. J. C}\ }\textbf
  {\bibinfo {volume} {71}},\ \bibinfo {pages} {1564} (\bibinfo {year}
  {2011})},\ \Eprint {http://arxiv.org/abs/1010.1216} {arXiv:1010.1216
  [hep-lat]} \BibitemShut {NoStop}%
\bibitem [{\citenamefont {Toublan}(1997)}]{Toublan:1997rr}%
  \BibitemOpen
  \bibfield  {author} {\bibinfo {author} {\bibfnamefont {D.}~\bibnamefont
  {Toublan}},\ }\href {\doibase 10.1103/PhysRevD.56.5629} {\bibfield  {journal}
  {\bibinfo  {journal} {Phys. Rev. D}\ }\textbf {\bibinfo {volume} {56}},\
  \bibinfo {pages} {5629} (\bibinfo {year} {1997})},\ \Eprint
  {http://arxiv.org/abs/hep-ph/9706273} {arXiv:hep-ph/9706273} \BibitemShut
  {NoStop}%
\bibitem [{\citenamefont {Andersen}(2012)}]{Andersen:2012dz}%
  \BibitemOpen
  \bibfield  {author} {\bibinfo {author} {\bibfnamefont {J.~O.}\ \bibnamefont
  {Andersen}},\ }\href {\doibase 10.1103/PhysRevD.86.025020} {\bibfield
  {journal} {\bibinfo  {journal} {Phys. Rev. D}\ }\textbf {\bibinfo {volume}
  {86}},\ \bibinfo {pages} {025020} (\bibinfo {year} {2012})},\ \Eprint
  {http://arxiv.org/abs/1202.2051} {arXiv:1202.2051 [hep-ph]} \BibitemShut
  {NoStop}%
\bibitem [{\citenamefont {Bernard}\ \emph {et~al.}(1987)\citenamefont
  {Bernard}, \citenamefont {Meissner},\ and\ \citenamefont
  {Zahed}}]{Bernard:1987ir}%
  \BibitemOpen
  \bibfield  {author} {\bibinfo {author} {\bibfnamefont {V.}~\bibnamefont
  {Bernard}}, \bibinfo {author} {\bibfnamefont {U.~G.}\ \bibnamefont
  {Meissner}}, \ and\ \bibinfo {author} {\bibfnamefont {I.}~\bibnamefont
  {Zahed}},\ }\href {\doibase 10.1103/PhysRevD.36.819} {\bibfield  {journal}
  {\bibinfo  {journal} {Phys. Rev. D}\ }\textbf {\bibinfo {volume} {36}},\
  \bibinfo {pages} {819} (\bibinfo {year} {1987})}\BibitemShut {NoStop}%
\bibitem [{\citenamefont {Sheng}\ \emph {et~al.}(2021)\citenamefont {Sheng},
  \citenamefont {Wang}, \citenamefont {Wang},\ and\ \citenamefont
  {Yu}}]{Sheng:2020hge}%
  \BibitemOpen
  \bibfield  {author} {\bibinfo {author} {\bibfnamefont {B.}~\bibnamefont
  {Sheng}}, \bibinfo {author} {\bibfnamefont {Y.}~\bibnamefont {Wang}},
  \bibinfo {author} {\bibfnamefont {X.}~\bibnamefont {Wang}}, \ and\ \bibinfo
  {author} {\bibfnamefont {L.}~\bibnamefont {Yu}},\ }\href {\doibase
  10.1103/PhysRevD.103.094001} {\bibfield  {journal} {\bibinfo  {journal}
  {Phys. Rev. D}\ }\textbf {\bibinfo {volume} {103}},\ \bibinfo {pages}
  {094001} (\bibinfo {year} {2021})},\ \Eprint
  {http://arxiv.org/abs/2010.05716} {arXiv:2010.05716 [hep-ph]} \BibitemShut
  {NoStop}%
\bibitem [{\citenamefont {Wergieluk}\ \emph {et~al.}(2013)\citenamefont
  {Wergieluk}, \citenamefont {Blaschke}, \citenamefont {Kalinovsky},\ and\
  \citenamefont {Friesen}}]{Wergieluk:2012gd}%
  \BibitemOpen
  \bibfield  {author} {\bibinfo {author} {\bibfnamefont {A.}~\bibnamefont
  {Wergieluk}}, \bibinfo {author} {\bibfnamefont {D.}~\bibnamefont {Blaschke}},
  \bibinfo {author} {\bibfnamefont {Y.~L.}\ \bibnamefont {Kalinovsky}}, \ and\
  \bibinfo {author} {\bibfnamefont {A.}~\bibnamefont {Friesen}},\ }\href
  {\doibase 10.1134/S1547477113070169} {\bibfield  {journal} {\bibinfo
  {journal} {Phys. Part. Nucl. Lett.}\ }\textbf {\bibinfo {volume} {10}},\
  \bibinfo {pages} {660} (\bibinfo {year} {2013})},\ \Eprint
  {http://arxiv.org/abs/1212.5245} {arXiv:1212.5245 [nucl-th]} \BibitemShut
  {NoStop}%
\bibitem [{\citenamefont {Blaschke}\ \emph {et~al.}(2018)\citenamefont
  {Blaschke}, \citenamefont {Dubinin}, \citenamefont {Ebert},\ and\
  \citenamefont {Friesen}}]{Blaschke:2017pvh}%
  \BibitemOpen
  \bibfield  {author} {\bibinfo {author} {\bibfnamefont {D.}~\bibnamefont
  {Blaschke}}, \bibinfo {author} {\bibfnamefont {A.}~\bibnamefont {Dubinin}},
  \bibinfo {author} {\bibfnamefont {D.}~\bibnamefont {Ebert}}, \ and\ \bibinfo
  {author} {\bibfnamefont {A.~V.}\ \bibnamefont {Friesen}},\ }\href {\doibase
  10.1134/S1547477118030056} {\bibfield  {journal} {\bibinfo  {journal} {Phys.
  Part. Nucl. Lett.}\ }\textbf {\bibinfo {volume} {15}},\ \bibinfo {pages}
  {230} (\bibinfo {year} {2018})},\ \Eprint {http://arxiv.org/abs/1712.09322}
  {arXiv:1712.09322 [hep-ph]} \BibitemShut {NoStop}%
\bibitem [{\citenamefont {Hatsuda}\ and\ \citenamefont
  {Kunihiro}(1987)}]{Hatsuda:1987kg}%
  \BibitemOpen
  \bibfield  {author} {\bibinfo {author} {\bibfnamefont {T.}~\bibnamefont
  {Hatsuda}}\ and\ \bibinfo {author} {\bibfnamefont {T.}~\bibnamefont
  {Kunihiro}},\ }\href {\doibase 10.1143/PTPS.91.284} {\bibfield  {journal}
  {\bibinfo  {journal} {Prog. Theor. Phys. Suppl.}\ }\textbf {\bibinfo {volume}
  {91}},\ \bibinfo {pages} {284} (\bibinfo {year} {1987})}\BibitemShut
  {NoStop}%
\bibitem [{\citenamefont {Andersen}(2007)}]{Andersen:2006ys}%
  \BibitemOpen
  \bibfield  {author} {\bibinfo {author} {\bibfnamefont {J.~O.}\ \bibnamefont
  {Andersen}},\ }\href {\doibase 10.1103/PhysRevD.75.065011} {\bibfield
  {journal} {\bibinfo  {journal} {Phys. Rev. D}\ }\textbf {\bibinfo {volume}
  {75}},\ \bibinfo {pages} {065011} (\bibinfo {year} {2007})},\ \Eprint
  {http://arxiv.org/abs/hep-ph/0609020} {arXiv:hep-ph/0609020} \BibitemShut
  {NoStop}%
\bibitem [{\citenamefont {Cui}\ and\ \citenamefont {Wu}(2013)}]{Cui:2013zha}%
  \BibitemOpen
  \bibfield  {author} {\bibinfo {author} {\bibfnamefont {L.-X.}\ \bibnamefont
  {Cui}}\ and\ \bibinfo {author} {\bibfnamefont {Y.-L.}\ \bibnamefont {Wu}},\
  }\href {\doibase 10.1142/S0217732313501320} {\bibfield  {journal} {\bibinfo
  {journal} {Mod. Phys. Lett. A}\ }\textbf {\bibinfo {volume} {28}},\ \bibinfo
  {pages} {1350132} (\bibinfo {year} {2013})},\ \Eprint
  {http://arxiv.org/abs/1302.4828} {arXiv:1302.4828 [hep-ph]} \BibitemShut
  {NoStop}%
\bibitem [{\citenamefont {Bartz}\ and\ \citenamefont
  {Jacobson}(2016)}]{Bartz:2016ufc}%
  \BibitemOpen
  \bibfield  {author} {\bibinfo {author} {\bibfnamefont {S.~P.}\ \bibnamefont
  {Bartz}}\ and\ \bibinfo {author} {\bibfnamefont {T.}~\bibnamefont
  {Jacobson}},\ }\href {\doibase 10.1103/PhysRevD.94.075022} {\bibfield
  {journal} {\bibinfo  {journal} {Phys. Rev. D}\ }\textbf {\bibinfo {volume}
  {94}},\ \bibinfo {pages} {075022} (\bibinfo {year} {2016})},\ \Eprint
  {http://arxiv.org/abs/1607.05751} {arXiv:1607.05751 [hep-ph]} \BibitemShut
  {NoStop}%
\bibitem [{\citenamefont {Roberts}(2008)}]{Roberts:2007ji}%
  \BibitemOpen
  \bibfield  {author} {\bibinfo {author} {\bibfnamefont {C.~D.}\ \bibnamefont
  {Roberts}},\ }\href {\doibase 10.1016/j.ppnp.2007.12.034} {\bibfield
  {journal} {\bibinfo  {journal} {Prog. Part. Nucl. Phys.}\ }\textbf {\bibinfo
  {volume} {61}},\ \bibinfo {pages} {50} (\bibinfo {year} {2008})},\ \Eprint
  {http://arxiv.org/abs/0712.0633} {arXiv:0712.0633 [nucl-th]} \BibitemShut
  {NoStop}%
\bibitem [{\citenamefont {Roberts}\ and\ \citenamefont
  {Williams}(1994)}]{Roberts:1994dr}%
  \BibitemOpen
  \bibfield  {author} {\bibinfo {author} {\bibfnamefont {C.~D.}\ \bibnamefont
  {Roberts}}\ and\ \bibinfo {author} {\bibfnamefont {A.~G.}\ \bibnamefont
  {Williams}},\ }\href {\doibase 10.1016/0146-6410(94)90049-3} {\bibfield
  {journal} {\bibinfo  {journal} {Prog. Part. Nucl. Phys.}\ }\textbf {\bibinfo
  {volume} {33}},\ \bibinfo {pages} {477} (\bibinfo {year} {1994})},\ \Eprint
  {http://arxiv.org/abs/hep-ph/9403224} {arXiv:hep-ph/9403224} \BibitemShut
  {NoStop}%
\bibitem [{\citenamefont {Xu}\ \emph {et~al.}(2019)\citenamefont {Xu},
  \citenamefont {Binosi}, \citenamefont {Cui}, \citenamefont {Li},
  \citenamefont {Roberts}, \citenamefont {Xu},\ and\ \citenamefont
  {Zong}}]{Xu:2019ilh}%
  \BibitemOpen
  \bibfield  {author} {\bibinfo {author} {\bibfnamefont {Y.-Z.}\ \bibnamefont
  {Xu}}, \bibinfo {author} {\bibfnamefont {D.}~\bibnamefont {Binosi}}, \bibinfo
  {author} {\bibfnamefont {Z.-F.}\ \bibnamefont {Cui}}, \bibinfo {author}
  {\bibfnamefont {B.-L.}\ \bibnamefont {Li}}, \bibinfo {author} {\bibfnamefont
  {C.~D.}\ \bibnamefont {Roberts}}, \bibinfo {author} {\bibfnamefont {S.-S.}\
  \bibnamefont {Xu}}, \ and\ \bibinfo {author} {\bibfnamefont {H.~S.}\
  \bibnamefont {Zong}},\ }\href {\doibase 10.1103/PhysRevD.100.114038}
  {\bibfield  {journal} {\bibinfo  {journal} {Phys. Rev. D}\ }\textbf {\bibinfo
  {volume} {100}},\ \bibinfo {pages} {114038} (\bibinfo {year} {2019})},\
  \Eprint {http://arxiv.org/abs/1911.05199} {arXiv:1911.05199 [nucl-th]}
  \BibitemShut {NoStop}%
\bibitem [{\citenamefont {Maris}\ and\ \citenamefont
  {Roberts}(2003)}]{Maris:2003vk}%
  \BibitemOpen
  \bibfield  {author} {\bibinfo {author} {\bibfnamefont {P.}~\bibnamefont
  {Maris}}\ and\ \bibinfo {author} {\bibfnamefont {C.~D.}\ \bibnamefont
  {Roberts}},\ }\href {\doibase 10.1142/S0218301303001326} {\bibfield
  {journal} {\bibinfo  {journal} {Int. J. Mod. Phys. E}\ }\textbf {\bibinfo
  {volume} {12}},\ \bibinfo {pages} {297} (\bibinfo {year} {2003})},\ \Eprint
  {http://arxiv.org/abs/nucl-th/0301049} {arXiv:nucl-th/0301049} \BibitemShut
  {NoStop}%
\bibitem [{\citenamefont {Maris}\ \emph {et~al.}(2001)\citenamefont {Maris},
  \citenamefont {Roberts}, \citenamefont {Schmidt},\ and\ \citenamefont
  {Tandy}}]{Maris:2000ig}%
  \BibitemOpen
  \bibfield  {author} {\bibinfo {author} {\bibfnamefont {P.}~\bibnamefont
  {Maris}}, \bibinfo {author} {\bibfnamefont {C.~D.}\ \bibnamefont {Roberts}},
  \bibinfo {author} {\bibfnamefont {S.~M.}\ \bibnamefont {Schmidt}}, \ and\
  \bibinfo {author} {\bibfnamefont {P.~C.}\ \bibnamefont {Tandy}},\ }\href
  {\doibase 10.1103/PhysRevC.63.025202} {\bibfield  {journal} {\bibinfo
  {journal} {Phys. Rev. C}\ }\textbf {\bibinfo {volume} {63}},\ \bibinfo
  {pages} {025202} (\bibinfo {year} {2001})},\ \Eprint
  {http://arxiv.org/abs/nucl-th/0001064} {arXiv:nucl-th/0001064} \BibitemShut
  {NoStop}%
\bibitem [{\citenamefont {Gao}\ and\ \citenamefont {Ding}(2020)}]{Gao:2020hwo}%
  \BibitemOpen
  \bibfield  {author} {\bibinfo {author} {\bibfnamefont {F.}~\bibnamefont
  {Gao}}\ and\ \bibinfo {author} {\bibfnamefont {M.}~\bibnamefont {Ding}},\
  }\href {\doibase 10.1140/epjc/s10052-020-08766-2} {\bibfield  {journal}
  {\bibinfo  {journal} {Eur. Phys. J. C}\ }\textbf {\bibinfo {volume} {80}},\
  \bibinfo {pages} {1171} (\bibinfo {year} {2020})},\ \Eprint
  {http://arxiv.org/abs/2006.05909} {arXiv:2006.05909 [hep-ph]} \BibitemShut
  {NoStop}%
\bibitem [{\citenamefont {Qin}(2015)}]{Qin:2013aaa}%
  \BibitemOpen
  \bibfield  {author} {\bibinfo {author} {\bibfnamefont {S.-x.}\ \bibnamefont
  {Qin}},\ }\href {\doibase 10.1016/j.physletb.2015.02.009} {\bibfield
  {journal} {\bibinfo  {journal} {Phys. Lett. B}\ }\textbf {\bibinfo {volume}
  {742}},\ \bibinfo {pages} {358} (\bibinfo {year} {2015})},\ \Eprint
  {http://arxiv.org/abs/1307.4587} {arXiv:1307.4587 [nucl-th]} \BibitemShut
  {NoStop}%
\bibitem [{\citenamefont {Chen}\ \emph {et~al.}(2020)\citenamefont {Chen},
  \citenamefont {Qin},\ and\ \citenamefont {Liu}}]{Chen:2020afc}%
  \BibitemOpen
  \bibfield  {author} {\bibinfo {author} {\bibfnamefont {L.-f.}\ \bibnamefont
  {Chen}}, \bibinfo {author} {\bibfnamefont {S.-X.}\ \bibnamefont {Qin}}, \
  and\ \bibinfo {author} {\bibfnamefont {Y.-x.}\ \bibnamefont {Liu}},\ }\href
  {\doibase 10.1103/PhysRevD.102.054015} {\bibfield  {journal} {\bibinfo
  {journal} {Phys. Rev. D}\ }\textbf {\bibinfo {volume} {102}},\ \bibinfo
  {pages} {054015} (\bibinfo {year} {2020})},\ \Eprint
  {http://arxiv.org/abs/2006.10582} {arXiv:2006.10582 [hep-ph]} \BibitemShut
  {NoStop}%
\bibitem [{\citenamefont {Yagi}\ \emph {et~al.}(2005)\citenamefont {Yagi},
  \citenamefont {Hatsuda},\ and\ \citenamefont {Miake}}]{Yagi:2005yb}%
  \BibitemOpen
  \bibfield  {author} {\bibinfo {author} {\bibfnamefont {K.}~\bibnamefont
  {Yagi}}, \bibinfo {author} {\bibfnamefont {T.}~\bibnamefont {Hatsuda}}, \
  and\ \bibinfo {author} {\bibfnamefont {Y.}~\bibnamefont {Miake}},\
  }\href@noop {} {\emph {\bibinfo {title} {{Quark-gluon plasma: From big bang
  to little bang}}}},\ Vol.~\bibinfo {volume} {23}\ (\bibinfo {year}
  {2005})\BibitemShut {NoStop}%
\bibitem [{\citenamefont {Delbourgo}\ and\ \citenamefont
  {Scadron}(1979)}]{Delbourgo:1979me}%
  \BibitemOpen
  \bibfield  {author} {\bibinfo {author} {\bibfnamefont {R.}~\bibnamefont
  {Delbourgo}}\ and\ \bibinfo {author} {\bibfnamefont {M.~D.}\ \bibnamefont
  {Scadron}},\ }\href {\doibase 10.1088/0305-4616/6/6/003} {\bibfield
  {journal} {\bibinfo  {journal} {J. Phys. G}\ }\textbf {\bibinfo {volume}
  {5}},\ \bibinfo {pages} {1621} (\bibinfo {year} {1979})},\ \bibinfo {note}
  {[Addendum: J.Phys.G 6, 649 (1980)]}\BibitemShut {NoStop}%
\bibitem [{\citenamefont {Bender}\ \emph {et~al.}(1996)\citenamefont {Bender},
  \citenamefont {Roberts},\ and\ \citenamefont {Von~Smekal}}]{Bender:1996bb}%
  \BibitemOpen
  \bibfield  {author} {\bibinfo {author} {\bibfnamefont {A.}~\bibnamefont
  {Bender}}, \bibinfo {author} {\bibfnamefont {C.~D.}\ \bibnamefont {Roberts}},
  \ and\ \bibinfo {author} {\bibfnamefont {L.}~\bibnamefont {Von~Smekal}},\
  }\href {\doibase 10.1016/0370-2693(96)00372-3} {\bibfield  {journal}
  {\bibinfo  {journal} {Phys. Lett. B}\ }\textbf {\bibinfo {volume} {380}},\
  \bibinfo {pages} {7} (\bibinfo {year} {1996})},\ \Eprint
  {http://arxiv.org/abs/nucl-th/9602012} {arXiv:nucl-th/9602012} \BibitemShut
  {NoStop}%
\bibitem [{\citenamefont {Munczek}(1995)}]{Munczek:1994zz}%
  \BibitemOpen
  \bibfield  {author} {\bibinfo {author} {\bibfnamefont {H.~J.}\ \bibnamefont
  {Munczek}},\ }\href {\doibase 10.1103/PhysRevD.52.4736} {\bibfield  {journal}
  {\bibinfo  {journal} {Phys. Rev. D}\ }\textbf {\bibinfo {volume} {52}},\
  \bibinfo {pages} {4736} (\bibinfo {year} {1995})},\ \Eprint
  {http://arxiv.org/abs/hep-th/9411239} {arXiv:hep-th/9411239} \BibitemShut
  {NoStop}%
\bibitem [{\citenamefont {Qin}\ \emph {et~al.}(2011{\natexlab{b}})\citenamefont
  {Qin}, \citenamefont {Chang}, \citenamefont {Liu}, \citenamefont {Roberts},\
  and\ \citenamefont {Wilson}}]{Qin:2011dd}%
  \BibitemOpen
  \bibfield  {author} {\bibinfo {author} {\bibfnamefont {S.-x.}\ \bibnamefont
  {Qin}}, \bibinfo {author} {\bibfnamefont {L.}~\bibnamefont {Chang}}, \bibinfo
  {author} {\bibfnamefont {Y.-x.}\ \bibnamefont {Liu}}, \bibinfo {author}
  {\bibfnamefont {C.~D.}\ \bibnamefont {Roberts}}, \ and\ \bibinfo {author}
  {\bibfnamefont {D.~J.}\ \bibnamefont {Wilson}},\ }\href {\doibase
  10.1103/PhysRevC.84.042202} {\bibfield  {journal} {\bibinfo  {journal} {Phys.
  Rev. C}\ }\textbf {\bibinfo {volume} {84}},\ \bibinfo {pages} {042202}
  (\bibinfo {year} {2011}{\natexlab{b}})},\ \Eprint
  {http://arxiv.org/abs/1108.0603} {arXiv:1108.0603 [nucl-th]} \BibitemShut
  {NoStop}%
\bibitem [{\citenamefont {Williams}\ \emph {et~al.}(2007)\citenamefont
  {Williams}, \citenamefont {Fischer},\ and\ \citenamefont
  {Pennington}}]{Williams:2006vva}%
  \BibitemOpen
  \bibfield  {author} {\bibinfo {author} {\bibfnamefont {R.}~\bibnamefont
  {Williams}}, \bibinfo {author} {\bibfnamefont {C.~S.}\ \bibnamefont
  {Fischer}}, \ and\ \bibinfo {author} {\bibfnamefont {M.~R.}\ \bibnamefont
  {Pennington}},\ }\href {\doibase 10.1016/j.physletb.2006.12.055} {\bibfield
  {journal} {\bibinfo  {journal} {Phys. Lett. B}\ }\textbf {\bibinfo {volume}
  {645}},\ \bibinfo {pages} {167} (\bibinfo {year} {2007})},\ \Eprint
  {http://arxiv.org/abs/hep-ph/0612061} {arXiv:hep-ph/0612061} \BibitemShut
  {NoStop}%
\bibitem [{\citenamefont {Fischer}(2009)}]{Fischer:2009wc}%
  \BibitemOpen
  \bibfield  {author} {\bibinfo {author} {\bibfnamefont {C.~S.}\ \bibnamefont
  {Fischer}},\ }\href {\doibase 10.1103/PhysRevLett.103.052003} {\bibfield
  {journal} {\bibinfo  {journal} {Phys. Rev. Lett.}\ }\textbf {\bibinfo
  {volume} {103}},\ \bibinfo {pages} {052003} (\bibinfo {year} {2009})},\
  \Eprint {http://arxiv.org/abs/0904.2700} {arXiv:0904.2700 [hep-ph]}
  \BibitemShut {NoStop}%
\bibitem [{\citenamefont {Blank}\ and\ \citenamefont
  {Krassnigg}(2010)}]{Blank:2010bz}%
  \BibitemOpen
  \bibfield  {author} {\bibinfo {author} {\bibfnamefont {M.}~\bibnamefont
  {Blank}}\ and\ \bibinfo {author} {\bibfnamefont {A.}~\bibnamefont
  {Krassnigg}},\ }\href {\doibase 10.1103/PhysRevD.82.034006} {\bibfield
  {journal} {\bibinfo  {journal} {Phys. Rev. D}\ }\textbf {\bibinfo {volume}
  {82}},\ \bibinfo {pages} {034006} (\bibinfo {year} {2010})},\ \Eprint
  {http://arxiv.org/abs/1004.5301} {arXiv:1004.5301 [hep-ph]} \BibitemShut
  {NoStop}%
\bibitem [{\citenamefont {Fischer}\ \emph {et~al.}(2014)\citenamefont
  {Fischer}, \citenamefont {Luecker},\ and\ \citenamefont
  {Welzbacher}}]{Fischer:2014ata}%
  \BibitemOpen
  \bibfield  {author} {\bibinfo {author} {\bibfnamefont {C.~S.}\ \bibnamefont
  {Fischer}}, \bibinfo {author} {\bibfnamefont {J.}~\bibnamefont {Luecker}}, \
  and\ \bibinfo {author} {\bibfnamefont {C.~A.}\ \bibnamefont {Welzbacher}},\
  }\href {\doibase 10.1103/PhysRevD.90.034022} {\bibfield  {journal} {\bibinfo
  {journal} {Phys. Rev. D}\ }\textbf {\bibinfo {volume} {90}},\ \bibinfo
  {pages} {034022} (\bibinfo {year} {2014})},\ \Eprint
  {http://arxiv.org/abs/1405.4762} {arXiv:1405.4762 [hep-ph]} \BibitemShut
  {NoStop}%
\bibitem [{\citenamefont {Contant}\ and\ \citenamefont
  {Huber}(2017)}]{Contant:2017gtz}%
  \BibitemOpen
  \bibfield  {author} {\bibinfo {author} {\bibfnamefont {R.}~\bibnamefont
  {Contant}}\ and\ \bibinfo {author} {\bibfnamefont {M.~Q.}\ \bibnamefont
  {Huber}},\ }\href {\doibase 10.1103/PhysRevD.96.074002} {\bibfield  {journal}
  {\bibinfo  {journal} {Phys. Rev. D}\ }\textbf {\bibinfo {volume} {96}},\
  \bibinfo {pages} {074002} (\bibinfo {year} {2017})},\ \Eprint
  {http://arxiv.org/abs/1706.00943} {arXiv:1706.00943 [hep-ph]} \BibitemShut
  {NoStop}%
\bibitem [{\citenamefont {Gao}\ and\ \citenamefont {Liu}(2016)}]{Gao:2016qkh}%
  \BibitemOpen
  \bibfield  {author} {\bibinfo {author} {\bibfnamefont {F.}~\bibnamefont
  {Gao}}\ and\ \bibinfo {author} {\bibfnamefont {Y.-x.}\ \bibnamefont {Liu}},\
  }\href {\doibase 10.1103/PhysRevD.94.076009} {\bibfield  {journal} {\bibinfo
  {journal} {Phys. Rev. D}\ }\textbf {\bibinfo {volume} {94}},\ \bibinfo
  {pages} {076009} (\bibinfo {year} {2016})},\ \Eprint
  {http://arxiv.org/abs/1607.01675} {arXiv:1607.01675 [hep-ph]} \BibitemShut
  {NoStop}%
\bibitem [{\citenamefont {Fischer}\ and\ \citenamefont
  {Luecker}(2013)}]{Fischer:2012vc}%
  \BibitemOpen
  \bibfield  {author} {\bibinfo {author} {\bibfnamefont {C.~S.}\ \bibnamefont
  {Fischer}}\ and\ \bibinfo {author} {\bibfnamefont {J.}~\bibnamefont
  {Luecker}},\ }\href {\doibase 10.1016/j.physletb.2012.11.054} {\bibfield
  {journal} {\bibinfo  {journal} {Phys. Lett. B}\ }\textbf {\bibinfo {volume}
  {718}},\ \bibinfo {pages} {1036} (\bibinfo {year} {2013})},\ \Eprint
  {http://arxiv.org/abs/1206.5191} {arXiv:1206.5191 [hep-ph]} \BibitemShut
  {NoStop}%
\bibitem [{\citenamefont {Maris}\ and\ \citenamefont
  {Roberts}(1997)}]{Maris:1997tm}%
  \BibitemOpen
  \bibfield  {author} {\bibinfo {author} {\bibfnamefont {P.}~\bibnamefont
  {Maris}}\ and\ \bibinfo {author} {\bibfnamefont {C.~D.}\ \bibnamefont
  {Roberts}},\ }\href {\doibase 10.1103/PhysRevC.56.3369} {\bibfield  {journal}
  {\bibinfo  {journal} {Phys. Rev. C}\ }\textbf {\bibinfo {volume} {56}},\
  \bibinfo {pages} {3369} (\bibinfo {year} {1997})},\ \Eprint
  {http://arxiv.org/abs/nucl-th/9708029} {arXiv:nucl-th/9708029} \BibitemShut
  {NoStop}%
\bibitem [{\citenamefont {Ding}\ \emph {et~al.}(2019)\citenamefont {Ding} \emph
  {et~al.}}]{HotQCD:2019xnw}%
  \BibitemOpen
  \bibfield  {author} {\bibinfo {author} {\bibfnamefont {H.~T.}\ \bibnamefont
  {Ding}} \emph {et~al.} (\bibinfo {collaboration} {HotQCD}),\ }\href {\doibase
  10.1103/PhysRevLett.123.062002} {\bibfield  {journal} {\bibinfo  {journal}
  {Phys. Rev. Lett.}\ }\textbf {\bibinfo {volume} {123}},\ \bibinfo {pages}
  {062002} (\bibinfo {year} {2019})},\ \Eprint
  {http://arxiv.org/abs/1903.04801} {arXiv:1903.04801 [hep-lat]} \BibitemShut
  {NoStop}%
\bibitem [{\citenamefont {Xu}\ \emph {et~al.}(2020)\citenamefont {Xu},
  \citenamefont {Shi}, \citenamefont {He},\ and\ \citenamefont
  {Zong}}]{Xu:2020loz}%
  \BibitemOpen
  \bibfield  {author} {\bibinfo {author} {\bibfnamefont {Y.-Z.}\ \bibnamefont
  {Xu}}, \bibinfo {author} {\bibfnamefont {C.}~\bibnamefont {Shi}}, \bibinfo
  {author} {\bibfnamefont {X.-T.}\ \bibnamefont {He}}, \ and\ \bibinfo {author}
  {\bibfnamefont {H.-S.}\ \bibnamefont {Zong}},\ }\href {\doibase
  10.1103/PhysRevD.102.114011} {\bibfield  {journal} {\bibinfo  {journal}
  {Phys. Rev. D}\ }\textbf {\bibinfo {volume} {102}},\ \bibinfo {pages}
  {114011} (\bibinfo {year} {2020})},\ \Eprint
  {http://arxiv.org/abs/2009.12035} {arXiv:2009.12035 [nucl-th]} \BibitemShut
  {NoStop}%
\bibitem [{\citenamefont {Steinbrecher}(2019)}]{Steinbrecher:2018phh}%
  \BibitemOpen
  \bibfield  {author} {\bibinfo {author} {\bibfnamefont {P.}~\bibnamefont
  {Steinbrecher}} (\bibinfo {collaboration} {HotQCD}),\ }\href {\doibase
  10.1016/j.nuclphysa.2018.08.025} {\bibfield  {journal} {\bibinfo  {journal}
  {Nucl. Phys. A}\ }\textbf {\bibinfo {volume} {982}},\ \bibinfo {pages} {847}
  (\bibinfo {year} {2019})},\ \Eprint {http://arxiv.org/abs/1807.05607}
  {arXiv:1807.05607 [hep-lat]} \BibitemShut {NoStop}%
\bibitem [{\citenamefont {Borsanyi}(2013)}]{Borsanyi:2012rr}%
  \BibitemOpen
  \bibfield  {author} {\bibinfo {author} {\bibfnamefont {S.}~\bibnamefont
  {Borsanyi}},\ }\href {\doibase 10.1016/j.nuclphysa.2013.01.072} {\bibfield
  {journal} {\bibinfo  {journal} {Nucl. Phys. A}\ }\textbf {\bibinfo {volume}
  {904-905}},\ \bibinfo {pages} {270c} (\bibinfo {year} {2013})},\ \Eprint
  {http://arxiv.org/abs/1210.6901} {arXiv:1210.6901 [hep-lat]} \BibitemShut
  {NoStop}%
\bibitem [{\citenamefont {Bazavov}\ \emph {et~al.}(2019)\citenamefont {Bazavov}
  \emph {et~al.}}]{HotQCD:2018pds}%
  \BibitemOpen
  \bibfield  {author} {\bibinfo {author} {\bibfnamefont {A.}~\bibnamefont
  {Bazavov}} \emph {et~al.} (\bibinfo {collaboration} {HotQCD}),\ }\href
  {\doibase 10.1016/j.physletb.2019.05.013} {\bibfield  {journal} {\bibinfo
  {journal} {Phys. Lett. B}\ }\textbf {\bibinfo {volume} {795}},\ \bibinfo
  {pages} {15} (\bibinfo {year} {2019})},\ \Eprint
  {http://arxiv.org/abs/1812.08235} {arXiv:1812.08235 [hep-lat]} \BibitemShut
  {NoStop}%
\bibitem [{\citenamefont {Qin}\ and\ \citenamefont
  {Rischke}(2014)}]{Qin:2014dqa}%
  \BibitemOpen
  \bibfield  {author} {\bibinfo {author} {\bibfnamefont {S.-x.}\ \bibnamefont
  {Qin}}\ and\ \bibinfo {author} {\bibfnamefont {D.~H.}\ \bibnamefont
  {Rischke}},\ }\href {\doibase 10.1016/j.physletb.2014.05.060} {\bibfield
  {journal} {\bibinfo  {journal} {Phys. Lett. B}\ }\textbf {\bibinfo {volume}
  {734}},\ \bibinfo {pages} {157} (\bibinfo {year} {2014})},\ \Eprint
  {http://arxiv.org/abs/1403.3025} {arXiv:1403.3025 [nucl-th]} \BibitemShut
  {NoStop}%
\bibitem [{\citenamefont {Karsch}\ \emph {et~al.}(2003)\citenamefont {Karsch},
  \citenamefont {Laermann}, \citenamefont {Petreczky},\ and\ \citenamefont
  {Stickan}}]{Karsch:2003wy}%
  \BibitemOpen
  \bibfield  {author} {\bibinfo {author} {\bibfnamefont {F.}~\bibnamefont
  {Karsch}}, \bibinfo {author} {\bibfnamefont {E.}~\bibnamefont {Laermann}},
  \bibinfo {author} {\bibfnamefont {P.}~\bibnamefont {Petreczky}}, \ and\
  \bibinfo {author} {\bibfnamefont {S.}~\bibnamefont {Stickan}},\ }\href
  {\doibase 10.1103/PhysRevD.68.014504} {\bibfield  {journal} {\bibinfo
  {journal} {Phys. Rev. D}\ }\textbf {\bibinfo {volume} {68}},\ \bibinfo
  {pages} {014504} (\bibinfo {year} {2003})},\ \Eprint
  {http://arxiv.org/abs/hep-lat/0303017} {arXiv:hep-lat/0303017} \BibitemShut
  {NoStop}%
\bibitem [{\citenamefont {Aarts}\ and\ \citenamefont
  {Martinez~Resco}(2005)}]{Aarts:2005hg}%
  \BibitemOpen
  \bibfield  {author} {\bibinfo {author} {\bibfnamefont {G.}~\bibnamefont
  {Aarts}}\ and\ \bibinfo {author} {\bibfnamefont {J.~M.}\ \bibnamefont
  {Martinez~Resco}},\ }\href {\doibase 10.1016/j.nuclphysb.2005.08.012}
  {\bibfield  {journal} {\bibinfo  {journal} {Nucl. Phys. B}\ }\textbf
  {\bibinfo {volume} {726}},\ \bibinfo {pages} {93} (\bibinfo {year} {2005})},\
  \Eprint {http://arxiv.org/abs/hep-lat/0507004} {arXiv:hep-lat/0507004}
  \BibitemShut {NoStop}%
\bibitem [{\citenamefont {Bryan}(1990)}]{Bryan:1990mx}%
  \BibitemOpen
  \bibfield  {author} {\bibinfo {author} {\bibfnamefont {R.}~\bibnamefont
  {Bryan}},\ }\href {\doibase 10.1007/BF02427376} {\bibfield  {journal}
  {\bibinfo  {journal} {European Biophysics Journal}\ }\textbf {\bibinfo
  {volume} {18}},\ \bibinfo {pages} {165} (\bibinfo {year} {1990})}\BibitemShut
  {NoStop}%
\bibitem [{\citenamefont {Nickel}(2007)}]{Nickel:2006mm}%
  \BibitemOpen
  \bibfield  {author} {\bibinfo {author} {\bibfnamefont {D.}~\bibnamefont
  {Nickel}},\ }\href {\doibase 10.1016/j.aop.2006.09.002} {\bibfield  {journal}
  {\bibinfo  {journal} {Annals Phys.}\ }\textbf {\bibinfo {volume} {322}},\
  \bibinfo {pages} {1949} (\bibinfo {year} {2007})},\ \Eprint
  {http://arxiv.org/abs/hep-ph/0607224} {arXiv:hep-ph/0607224} \BibitemShut
  {NoStop}%
\bibitem [{\citenamefont {Asakawa}\ \emph {et~al.}(2001)\citenamefont
  {Asakawa}, \citenamefont {Hatsuda},\ and\ \citenamefont
  {Nakahara}}]{Asakawa:2000tr}%
  \BibitemOpen
  \bibfield  {author} {\bibinfo {author} {\bibfnamefont {M.}~\bibnamefont
  {Asakawa}}, \bibinfo {author} {\bibfnamefont {T.}~\bibnamefont {Hatsuda}}, \
  and\ \bibinfo {author} {\bibfnamefont {Y.}~\bibnamefont {Nakahara}},\ }\href
  {\doibase 10.1016/S0146-6410(01)00150-8} {\bibfield  {journal} {\bibinfo
  {journal} {Prog. Part. Nucl. Phys.}\ }\textbf {\bibinfo {volume} {46}},\
  \bibinfo {pages} {459} (\bibinfo {year} {2001})},\ \Eprint
  {http://arxiv.org/abs/hep-lat/0011040} {arXiv:hep-lat/0011040} \BibitemShut
  {NoStop}%
\bibitem [{\citenamefont {Mueller}\ \emph {et~al.}(2010)\citenamefont
  {Mueller}, \citenamefont {Fischer},\ and\ \citenamefont
  {Nickel}}]{Mueller:2010ah}%
  \BibitemOpen
  \bibfield  {author} {\bibinfo {author} {\bibfnamefont {J.~A.}\ \bibnamefont
  {Mueller}}, \bibinfo {author} {\bibfnamefont {C.~S.}\ \bibnamefont
  {Fischer}}, \ and\ \bibinfo {author} {\bibfnamefont {D.}~\bibnamefont
  {Nickel}},\ }\href {\doibase 10.1140/epjc/s10052-010-1499-8} {\bibfield
  {journal} {\bibinfo  {journal} {Eur. Phys. J. C}\ }\textbf {\bibinfo {volume}
  {70}},\ \bibinfo {pages} {1037} (\bibinfo {year} {2010})},\ \Eprint
  {http://arxiv.org/abs/1009.3762} {arXiv:1009.3762 [hep-ph]} \BibitemShut
  {NoStop}%
\bibitem [{\citenamefont {Qin}\ and\ \citenamefont
  {Rischke}(2013)}]{Qin:2013ufa}%
  \BibitemOpen
  \bibfield  {author} {\bibinfo {author} {\bibfnamefont {S.-x.}\ \bibnamefont
  {Qin}}\ and\ \bibinfo {author} {\bibfnamefont {D.~H.}\ \bibnamefont
  {Rischke}},\ }\href {\doibase 10.1103/PhysRevD.88.056007} {\bibfield
  {journal} {\bibinfo  {journal} {Phys. Rev. D}\ }\textbf {\bibinfo {volume}
  {88}},\ \bibinfo {pages} {056007} (\bibinfo {year} {2013})},\ \Eprint
  {http://arxiv.org/abs/1304.6547} {arXiv:1304.6547 [nucl-th]} \BibitemShut
  {NoStop}%
\bibitem [{\citenamefont {Qin}\ and\ \citenamefont
  {Roberts}(2021)}]{Qin:2020jig}%
  \BibitemOpen
  \bibfield  {author} {\bibinfo {author} {\bibfnamefont {S.-X.}\ \bibnamefont
  {Qin}}\ and\ \bibinfo {author} {\bibfnamefont {C.~D.}\ \bibnamefont
  {Roberts}},\ }\href {\doibase 10.1088/0256-307X/38/7/071201} {\bibfield
  {journal} {\bibinfo  {journal} {Chin. Phys. Lett.}\ }\textbf {\bibinfo
  {volume} {38}},\ \bibinfo {pages} {071201} (\bibinfo {year} {2021})},\
  \Eprint {http://arxiv.org/abs/2009.13637} {arXiv:2009.13637 [hep-ph]}
  \BibitemShut {NoStop}%
\end{thebibliography}
\end{document}